\documentclass{article}

\usepackage{arxiv}

\usepackage[utf8]{inputenc} 
\usepackage[T1]{fontenc}    
\usepackage{hyperref}       
\usepackage{url}            
\usepackage{booktabs,tabularx,longtable,multirow,array}
\usepackage{amsfonts,amsmath,amssymb}
\usepackage{nicefrac}       
\usepackage{microtype}      
\usepackage[numbers,sort&compress]{natbib}
\usepackage{doi}
\usepackage[table]{xcolor}
\usepackage{tikz}
\usetikzlibrary{positioning,arrows.meta,fit,calc,shapes.geometric}
\usepackage{listings}
\usepackage{caption}
\usepackage{enumitem}
\usepackage{fontawesome5}
\usepackage[section]{placeins}   
\usepackage[capitalise,noabbrev]{cleveref}

\definecolor{sblue}{HTML}{2A78D6}
\definecolor{sorange}{HTML}{EB6834}
\definecolor{saqua}{HTML}{1BAF7A}
\definecolor{sgray}{HTML}{52514E}
\definecolor{sline}{HTML}{E4E3DF}
\definecolor{sred}{HTML}{E34948}
\definecolor{sviolet}{HTML}{4A3AA7}

\setlist{itemsep=1pt, topsep=3pt}

\newcommand{\evt}[1]{\texttt{#1}}   
\DeclareRobustCommand{\fire}{{\color{sorange}\faFire}}
\DeclareRobustCommand{\ice}{{\color{sblue}\faSnowflake[regular]}}

\title{Speech Event Detector: Timestamped Detection of Non-Verbal Vocalizations and Background Sounds for Rich Speech Transcription}

\renewcommand{\shorttitle}{Speech Event Detector}

\author{
\begin{tabular}{c@{\hspace{2cm}}c}
\href{https://orcid.org/0000-0003-3050-6478}
{\includegraphics[scale=0.06]{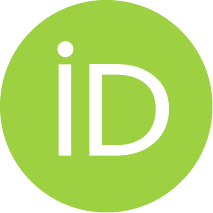}\hspace{1mm}Ayush Kumar Bar}
&
\href{https://orcid.org/0000-0001-8611-7004}
{\includegraphics[scale=0.06]{orcid.pdf}\hspace{1mm}Promila Ghosh}
\end{tabular}
\\[0.5em]
\texttt{\href{https://github.com/theboringai-work/surround-sound}
{\faGithub\enspace github.com/theboringai-work/surround-sound}}
\\[0.3em]
\texttt{\href{https://huggingface.co/theboringai-work/surround-sound}
{\faSmile[regular]\enspace huggingface.co/theboringai-work/surround-sound}}
}

\date{}

\hypersetup{
  pdftitle={Speech Event Detector: Timestamped Detection of Non-Verbal Vocalizations and Background Sounds for Rich Speech Transcription},
  pdfauthor={Ayush Kumar Bar},
  pdfkeywords={sound event detection, non-verbal vocalizations, rich transcription, synthetic mixtures, forced alignment},
  hidelinks
}

\begin{document}
\maketitle

\begin{abstract}
Automatic speech recognition produces words but drops the non-verbal content of a recording: the breaths, coughs, sighs and laughter of the speaker, and the horns, music or barking around them. We describe \emph{Speech Event Detector}, a 5.8\,M-parameter convolutional--Transformer model that outputs, every 40\,ms, probabilities for 54 event classes (20 speaker vocalizations and 34 background sounds), together with the post-processing that turns these probabilities into timestamped events and a procedure that merges the events into a Whisper transcript. The model is trained on material drawn from seven public datasets whose labels come in incompatible forms: single detected regions, free-text tags with timestamps, temporally strong ontology labels, symbols embedded in transcripts, isolated recordings and environment recordings. We unify them with rule-based label mapping, recover timestamps for transcript-embedded labels by CTC forced alignment (6{,}360 of 8{,}620 sounds located), and combine about 168\,h of real annotated training recordings with synthetic mixtures and augmentation using real background noise. Per-class thresholds for two operating profiles are tuned on real held-out recordings, and their performance is estimated by two-fold cross-validation. On whole held-out recordings from four datasets, over the 48 classes with at least ten held-out events, the sensitive profile finds on average 49\,\% of a class's events (67\,\% of all 9{,}367 events) at a macro precision of 0.30, and the balanced profile reaches a macro recall of 0.37 at a precision of 0.36 (macro event F1 0.32 and 0.33). The detector runs about 39 times faster than real time on eight CPU threads. We document the evaluation's limitations, in particular that thresholds were tuned on held-out data overlapping the evaluation clips and that no standard benchmark was run, and the non-commercial terms of three training sources.
\end{abstract}


\section{Introduction}
\label{sec:intro}

Transcripts produced by modern speech recognizers such as Whisper~\citep{radford2023whisper} are optimized for the words. Much of what a listener hears is not words: a speaker's sigh before an answer, laughter, a cough that interrupts a sentence, or a horn, music or a barking dog in the background. These events carry meaning for conversation analysis, for captioning, for curating expressive speech corpora, and for filtering training data. Recent corpora for expressive and non-verbal speech, such as NonverbalTTS~\citep{borisov2025nonverbaltts} and NonVerbalSpeech-38K~\citep{nvs38k}, show the demand for transcripts that include such events.

Sound event detection (SED), the task of recognizing sound classes together with their onset and offset times, is well studied~\citep{mesaros2021tutorial,cakir2017crnn}, but most systems target environmental or domestic sounds~\citep{turpault2019desed}, and general-purpose audio models are usually trained for clip-level tagging~\citep{kong2020panns,chen2023beats,gong2023whisperat}. Detecting the non-verbal sounds that occur \emph{inside} speech, at a time resolution useful for placing a tag between two words, raises three practical problems:
\begin{enumerate}
\item \textbf{Fragmented, heterogeneous labels.} Public data with timestamps for these sounds is scarce and spread across corpora that label them in incompatible ways: one detected region per clip, free-text tags with timestamps, ontology labels, symbols embedded in transcripts without timing, or whole recordings of a single sound or environment.
\item \textbf{Events overlap speech.} Vocalizations are short, often overlap words, and resemble speech acoustically; background events are long and loosely bounded.
\item \textbf{Operating point.} A transcript tag is useful only if false tags are rare enough to be ignored, while in our own early use missed events were the main complaint; a single global threshold serves neither need.
\end{enumerate}

\begin{figure*}[!htbp]
\centering
\resizebox{\textwidth}{!}{\input{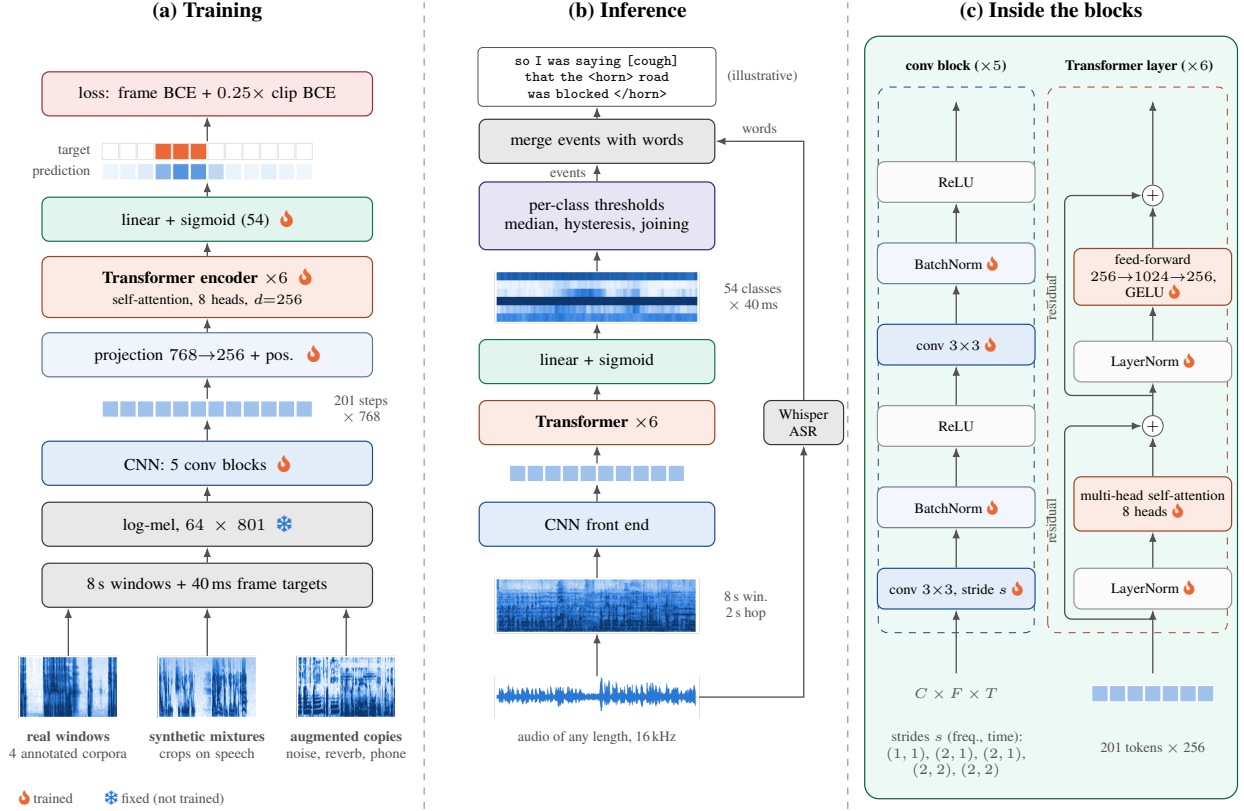}}
\caption{System overview. \textbf{(a)} Training: real annotated windows, synthetic mixtures and augmented copies (thumbnails are log-mel spectrograms of actual training examples) become 8\,s windows with 40\,ms frame targets; the network is trained with a frame-level loss plus a clip-level term. \textbf{(b)} Inference: audio of any length is processed in 8\,s windows (thumbnails show a held-out clip, its spectrogram and the resulting class probabilities); per-class thresholds and post-processing give timestamped events, which are merged with the word timestamps of a Whisper transcript. \textbf{(c)} The two repeated building blocks. \protect\faFire\ marks trained parts, \protect\faSnowflake[regular]\ fixed ones. Section references: data \S\ref{sec:data}, model \S\ref{sec:model}, training \S\ref{sec:training}, inference \S\ref{sec:inference}.}
\label{fig:pipeline}
\end{figure*}

This report describes a practical system that addresses these problems with modest compute (one NVIDIA T4 GPU), and documents it in detail (exact reproduction also requires an unreleased preliminary checkpoint; \cref{app:repro}). Our contributions are:
\begin{itemize}
\item a \textbf{label-unification pipeline} that maps seven public sources into one taxonomy of 54 timestamped classes (\cref{sec:task,sec:data}), including the recovery of timestamps for transcript-embedded labels by CTC forced alignment (\cref{sec:tts});
\item a \textbf{training recipe} that combines real annotated recordings with synthetic mixtures whose event positions are known by construction, augments both with real background noise, and selects the model by average precision on real held-out audio, under a memory budget that fits a single GPU machine (\cref{sec:model,sec:training});
\item \textbf{per-class calibration} into two operating profiles by cross-validation on real held-out audio, and a \textbf{transcript merge} that inserts tags without ever reordering the recognizer's words (\cref{sec:inference});
\item an \textbf{evaluation on held-out real recordings} from four datasets with an honest account of its limitations (\cref{sec:eval,sec:ethics}).
\end{itemize}
\Cref{fig:pipeline} gives an overview. Code is available at \url{https://github.com/theboringai-work/surround-sound} and pre-trained weights are hosted at \url{https://huggingface.co/theboringai-work/surround-sound}; the weights are restricted to non-commercial use by the terms of the training data (\cref{sec:ethics}).

\section{Related work}
\label{sec:related}

\paragraph{Sound event detection.} SED assigns class activity to short frames and derives events with onsets and offsets from them~\citep{mesaros2021tutorial}. Convolutional--recurrent networks that combine spectro-temporal convolution with sequence modelling became a standard design~\citep{cakir2017crnn,turpault2019desed}; our model follows the same pattern with a Transformer encoder~\citep{vaswani2017attention} in place of the recurrent layers. Evaluation uses segment- and event-based metrics with collars~\citep{mesaros2016metrics} or threshold-free measures such as the polyphonic sound detection score~\citep{bilen2020psds}. Our event-level evaluation is a lenient overlap-based variant (\cref{sec:protocol}).

\paragraph{Synthetic mixtures.} Because strongly labelled recordings are expensive, SED systems are routinely trained on synthetic soundscapes in which isolated events are mixed into backgrounds at known times, for example with Scaper~\citep{salamon2017scaper} in the DESED setup~\citep{turpault2019desed}. We use the same idea with speech as the background, because our target events occur inside speech.

\paragraph{Large-scale audio tagging.} AudioSet~\citep{gemmeke2017audioset} provides an ontology of sound classes and millions of weakly labelled clips; a subset was later annotated with temporally strong labels, which improve event classification~\citep{hershey2021benefit}. Pretrained taggers such as PANNs~\citep{kong2020panns} and BEATs~\citep{chen2023beats}, and datasets such as FSD50K~\citep{fonseca2022fsd50k} and ESC-50~\citep{piczak2015esc}, target general sounds. Whisper-AT~\citep{gong2023whisperat} shows that the representations of a speech recognizer can support clip-level audio tagging with little extra cost. Self-supervised speech models such as WavLM~\citep{chen2022wavlm} capture paralinguistic information. Our model is trained from log-mel features without such a backbone, which keeps it small; we return to this choice in \cref{sec:ethics}.

\paragraph{Non-verbal vocalizations.} Laughter detection in noisy recordings has been studied with precise segmentations on AudioSet~\citep{gillick2021laughter}. Recent corpora annotate several vocalization types: NonverbalTTS~\citep{borisov2025nonverbaltts} marks ten types inside English transcripts derived from VoxCeleb~\citep{nagrani2017voxceleb} and Expresso~\citep{nguyen2023expresso}; NonVerbalSpeech-38K~\citep{nvs38k} provides one detected region for each of 38{,}718 mostly Mandarin clips; the Deeply Nonverbal Vocalization dataset~\citep{deeply2021} contains isolated recordings of 16 types; and the Vaani noise-event release~\citep{vaani_noise,shukla2026vaaninoise,pulikodan2026vaani} labels both vocal and background events in Indic speech across 58 Indian languages with timestamps. Our contribution lies in combining these sources into one timestamped detector and in calibrating and evaluating it, rather than in the network architecture.

\section{Task and label taxonomy}
\label{sec:task}

Given audio $x$ resampled to 16\,kHz mono, the model outputs a probability $p_{t,c}\in[0,1]$ for every class $c$ and every frame $t$ of 40\,ms. Post-processing (\cref{sec:postproc}) turns the probability tracks into events $(c, t_\text{start}, t_\text{end}, \bar p)$, where $\bar p$ is the mean frame probability over the event.

\paragraph{Event list.} We listed 55 events in two groups. The \emph{speaker sounds} are sounds a talker makes: \evt{breath}, \evt{sigh}, \evt{sniff}, \evt{laugh}, \evt{cough}, \evt{throat\_clear}, \evt{cry}, \evt{gasp}, \evt{yawn}, \evt{snore}, \evt{sneeze}, \evt{nose\_blow}, \evt{lip\_smack}, \evt{groan}, \evt{scream}, \evt{whisper}, \evt{shout}, \evt{chew}, \evt{burp}, \evt{hiccup} and \evt{click}. The \emph{background sounds} are \evt{horn}, \evt{vehicle}, \evt{siren}, \evt{bird}, \evt{insect}, \evt{dog\_bark}, \evt{animal}, \evt{child\_voice}, \evt{baby\_cry}, \evt{music}, \evt{tv\_speaker}, \evt{machine}, \evt{fan}, \evt{phone\_ring}, \evt{phone\_vibrate}, \evt{beep}, \evt{bell}, \evt{whistle}, \evt{buzzer}, \evt{applause}, \evt{cheer}, \evt{crowd}, \evt{footsteps}, \evt{wind}, \evt{rain}, \evt{thunder}, \evt{water}, \evt{knock}, \evt{door}, \evt{dishes}, \evt{glass\_break}, \evt{gunshot}, \evt{explosion} and \evt{construction}. A class is trained only if the sources provide at least 80 annotated events for it. This excludes \evt{click}, which has 47 (all from OpenSLR\,99), and leaves \textbf{54 classes: 20 speaker sounds and 34 background sounds}. \Cref{tab:counts} in the appendix lists the annotated events per class and source.

\paragraph{Rendering.} In a transcript, a speaker sound is written inline as \texttt{[tag]} at the position where it starts. A background sound opens with \texttt{<tag>} at its onset and, if words fall inside it and it ends at least 0.3\,s before the end of the transcript's last word, closes with \texttt{</tag>} after the last word it overlaps; otherwise it is written as a lone \texttt{<tag>}. For example: \texttt{so I was saying [cough] that the <horn> road was blocked </horn> anyway}.

\section{Data}
\label{sec:data}

\begin{table*}[!htbp]
\centering\small
\caption{Training and evaluation sources. Sizes are those of the full releases as indexed for this work. Licenses are quoted from each source's own page; see \cref{sec:ethics} for their consequences.}
\label{tab:datasets}
\begin{tabularx}{\textwidth}{@{}p{2.6cm}p{3.0cm}X p{2.2cm}p{2.3cm}@{}}
\toprule
Source & Size & Labels and use & Held-out part & License \\
\midrule
NonVerbalSpeech-38K \citep{nvs38k} & 38{,}718 clips, 130.6\,h; 36{,}925 Mandarin, 1{,}793 English & one non-verbal event per clip with a detected region; event crops, speech beds, real windows & 10\,\% by stable hash (3{,}828 clips) & CC BY-NC 4.0 \\
Vaani noise events \citep{vaani_noise,shukla2026vaaninoise,pulikodan2026vaani} & 90{,}637 clips, 154.6\,h; 58 Indian languages + English, 162 districts & 319 free-text noise tags with start/end (11{,}111 verified, 61{,}642 unverified, 17{,}884 without timestamps); crops, real windows & 14 districts, verified clips only & CC BY 4.0 \\
NonverbalTTS \citep{borisov2025nonverbaltts} & 6{,}256 clips, 17.6\,h, English (VoxCeleb, Expresso) & 8{,}620 sounds marked inside transcripts, no timestamps; forced alignment (\cref{sec:tts}); real windows, crops, speech beds & official dev + test (405 clips) & annotations CC BY-NC-SA 4.0; audio per source corpora \\
AudioSet-Strong \citep{hershey2021benefit,audioset_strong_mirror} & 93{,}609 10\,s clips (79{,}564 train, 14{,}045 eval), 260\,h & 455 event names with start/end (220 mapped); streamed subset; real windows, crops & official evaluation split & not stated on the mirror \\
OpenSLR 99 \citep{deeply2021} & 727 clips, 410 speakers, 16 classes & isolated phone-recorded vocalizations; trimmed crops & 15\,\% of speakers & CC BY-NC-ND 4.0 \\
DNC \citep{zequeira2021dnc} & 4{,}377 home recordings & noise type; ambient crops (TV/radio, music, cars, appliances) and a background-noise pool & 10\,\% of sessions & MIT \\
DEMAND \citep{thiemann2013demand} & 18 environments, 16 channels & environment; channels 1 and 9; ambient crops (transport, washing room) and noise pool & last 20\,\% of each recording & CC BY 4.0 (record text also states CC BY-SA 3.0) \\
\bottomrule
\end{tabularx}
\end{table*}

\Cref{tab:datasets} lists the seven sources. All statistics in this section were recomputed from the cached indices and audio banks of the training run (the released \texttt{data/stats.json}).

\subsection{Label unification}
\label{sec:labels}
Every source is mapped to the taxonomy of \cref{sec:task}; the complete rules are reproduced in \cref{app:rules}.
\begin{itemize}
\item \textbf{NonVerbalSpeech-38K} (NVS-38K) labels are mapped one to one; \evt{laughing} becomes \evt{laugh}, \evt{throatclearing} becomes \evt{throat\_clear}, and so on.
\item \textbf{Vaani} has 319 distinct free-text tags such as \texttt{<horn>}, \texttt{<honking>} or \texttt{<vehicle horn>}. An ordered list of regular expressions (first match wins) maps each tag to a class, to \emph{ignore} for tags that denote speech (e.g.\ \texttt{<child talking>}), or to \emph{drop} for the rest. Of the 106{,}892 tag occurrences, 99{,}320 map to a class, 7{,}502 are ignored and 70 are dropped; a clip that contains a dropped tag is not used as a real annotated window, since one of its sounds would be unlabelled.
\item \textbf{AudioSet-Strong} uses 455 event names; a second ordered rule list maps 220 of them to our classes. The rest (speech, generic impacts and other classes outside the taxonomy) are not targets.
\item \textbf{NonverbalTTS} marks each sound with one emoji inside the transcript; the ten emoji map to nine classes, with grunting merged into \evt{groan}.
\item \textbf{OpenSLR\,99} folder names map to classes (panting $\rightarrow$ \evt{breath}, lip-popping $\rightarrow$ \evt{lip\_smack}, moaning $\rightarrow$ \evt{groan}); teeth chattering and grinding are not used.
\item \textbf{DNC} recordings of TV and radio map to \evt{tv\_speaker}, music to \evt{music}, cars to \evt{vehicle}, and appliances (washing machine, water heater, dishwasher, coffee machine, vacuum cleaner, microwave) to \evt{machine}. \textbf{DEMAND} traffic, bus, car and metro map to \evt{vehicle} and the washing room to \evt{machine}. The remaining DNC recordings (quiet rooms, people) and DEMAND environments (kitchen, living room, field, park, river, hallway, meeting, office, cafeteria, restaurant, station, caf\'{e}, square) become an unlabelled background-noise pool used only for augmentation.
\end{itemize}

\subsection{Timestamps for NonverbalTTS by forced alignment}
\label{sec:tts}
NonverbalTTS gives the position of each sound in the transcript (between two words) but no times. We recover times in three steps.
\begin{enumerate}
\item \textbf{Word alignment.} The normalized words are aligned to the audio with the MMS forced-alignment model~\citep{pratap2024mms} as packaged in TorchAudio~\citep{yang2022torchaudio,hwang2023torchaudio}, using a Viterbi search over the CTC~\citep{graves2006ctc} emissions implemented in plain PyTorch~\citep{paszke2019pytorch}; it is written to reproduce TorchAudio's own aligner. A wildcard token at both ends absorbs untranscribed speech. Clips whose mean alignment probability is below 0.35 are skipped.
\item \textbf{Event placement.} A sound is searched for in the pause between the word before and the word after its symbol, shrunk by 60\,ms at each side. In 20\,ms frames, the event is the run of frames more than 4\,dB above the clip's noise floor (its 10th-percentile frame energy) with the highest total energy, joining runs less than 100\,ms apart and padding by 40\,ms. If no frame rises above the floor, the whole pause is used when its length is plausible. Each class has a maximum duration (from 1.0\,s for \evt{sniff} to 4.0\,s for \evt{laugh}); longer spans are cut around the loudest frame. Pauses shorter than about 0.2\,s, as when laughing while talking, are not located.
\item \textbf{Use.} Located sounds become event crops. Clips in which every sound was located, and clips without any sound, become real annotated windows; clips without any sound (and at least 2\,s long) also become English speech beds. Clips longer than 30\,s are not aligned.
\end{enumerate}
Of the 8{,}620 annotated sounds, \textbf{6{,}360 (74\,\%) were located}: 4{,}546 breaths, 879 laughs, 327 sniffs, 224 coughs, 187 throat clearings, 107 sighs, 69 groans, 13 sneezes and 8 snores. 4{,}549 clips became real windows (3{,}297 with every sound located and 1{,}252 without sounds), and 1{,}243 of the latter also became speech beds. These timestamps were not manually verified; they inherit any error of the transcript symbol position, which the dataset's authors derived with an automatic detector and forced alignment before human validation~\citep{borisov2025nonverbaltts}.

\subsection{Splits}
\label{sec:splits}
Validation clips are held out from training (\cref{tab:datasets}), with two small exceptions: the babble augmentation draws on speech beds of both splits (including 990 NVS-38K validation beds), and the unlabelled background-noise pool is not split. NVS-38K uses a 10\,\% split by a CRC-32 hash of the clip's file, row group and row; Vaani holds out 14 of its 162 districts and validates only on their verified clips, discarding the unverified ones from those districts entirely. NonverbalTTS and AudioSet-Strong use their official development/test and evaluation splits, OpenSLR\,99 holds out 15\,\% of speakers, DNC 10\,\% of recording sessions, and DEMAND the last 20\,\% of each labelled recording.

\subsection{Audio banks}
\label{sec:banks}
\begin{figure}[!htbp]
\centering
\includegraphics[width=\linewidth]{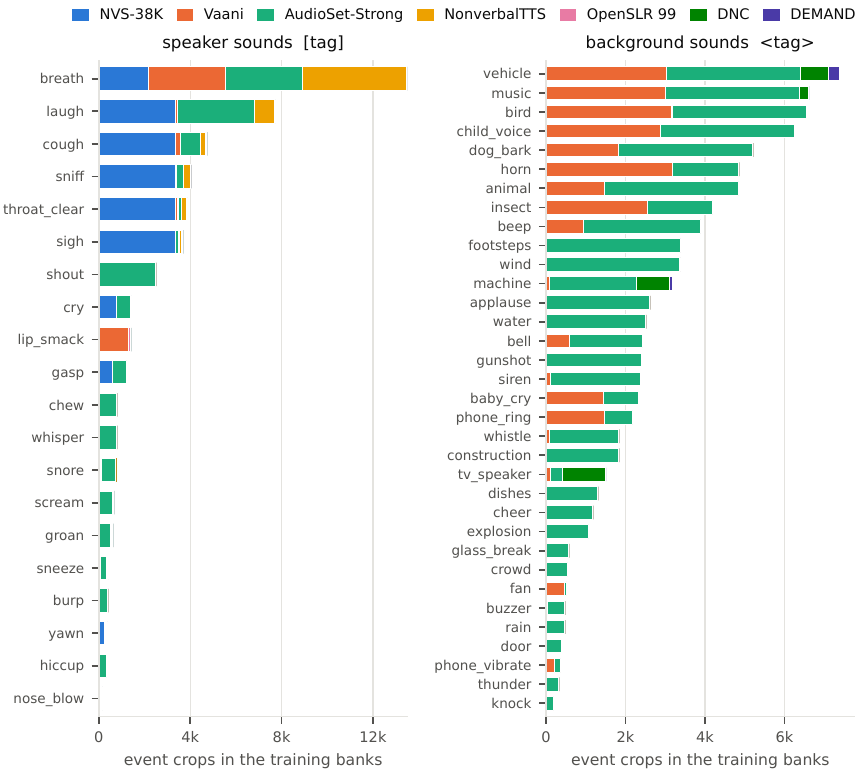}
\caption{Event crops per class in the audio banks (training and validation), by source. AudioSet-Strong supplies most background classes and most crops of the rarer speaker classes; NVS-38K and NonverbalTTS dominate \evt{breath}, \evt{cough}, \evt{laugh}, \evt{sigh}, \evt{sniff} and \evt{throat\_clear}; Vaani dominates \evt{lip\_smack} and \evt{nose\_blow}; DNC is the main source of \evt{tv\_speaker}. Exact counts are in the released \texttt{data/event\_crops\_by\_source.csv}.}
\label{fig:crops}
\end{figure}

All audio is decoded, resampled to 16\,kHz and stored as int16 in memory-mapped files, so that RAM use does not grow with the data. For the parquet-based sources, a metadata index records the row group of every clip; all NonverbalTTS clips are processed, while for the others a greedy pass then selects the row groups that best fill per-class quotas, weighting each class by the inverse of its available count so that rare classes count as much as common ones. AudioSet-Strong (121\,GB) is streamed rather than downloaded: up to 500 training and 60 evaluation row groups of 72 clips each are read, and only the chosen clips are stored, 10\,\% of them without any target event. Event crops keep a 30\,ms margin around the labelled span. For NVS-38K, Vaani, AudioSet-Strong, DNC and DEMAND about 3{,}000 training crops per class and source are kept (the cap is checked per clip, so AudioSet-Strong slightly exceeds it for a few classes); all located NonverbalTTS sounds are kept. \Cref{tab:banks} summarizes the banks and \cref{fig:crops} shows how the crops of each class are distributed over the sources.

\begin{table}[!htbp]
\centering\small
\caption{Audio banks (16\,kHz int16, memory-mapped) and the cached feature pools of 8\,s examples used for training and model selection.}
\label{tab:banks}
\begin{tabular}{@{}lrr@{}}
\toprule
Audio bank & Items & Hours \\
\midrule
real windows: NVS-38K, Vaani & 37{,}434 & 80.6 \\
real windows: AudioSet-Strong & 33{,}750 & 92.5 \\
real windows: NonverbalTTS & 4{,}549 & 11.8 \\
event crops: NVS-38K, Vaani & 52{,}654 & 25.2 \\
event crops: AudioSet-Strong & 75{,}343 & 25.9 \\
event crops: NonverbalTTS & 6{,}360 & 1.0 \\
event crops: OpenSLR 99 & 591 & 0.4 \\
ambient crops: DNC, DEMAND & 3{,}234 & 12.3 \\
speech beds: NVS-38K & 8{,}862 & 18.0 \\
speech beds: NonverbalTTS & 1{,}243 & 2.6 \\
background-noise pool: DNC, DEMAND & 2{,}279 & 6.3 \\
\midrule
Feature pool (8\,s windows) & Examples & Events \\
\midrule
synthetic mixes (clean + 1 augmented copy) & 60{,}000 & 83{,}344 \\
real windows NVS-38K (clean + 2 copies) & 27{,}000 & 26{,}110 \\
real windows Vaani (clean + 2 copies) & 27{,}000 & 65{,}179 \\
real windows NonverbalTTS (clean + 2 copies) & 21{,}000 & 21{,}749 \\
real windows AudioSet-Strong (clean + 2 copies) & 36{,}000 & 142{,}569 \\
\textbf{training total} & \textbf{171{,}000} & \\
\midrule
validation: NVS-38K & 600 & 584 \\
validation: Vaani & 800 & 1{,}777 \\
validation: NonverbalTTS & 400 & 558 \\
validation: AudioSet-Strong & 800 & 3{,}729 \\
validation: noisy (augmented real) & 600 & 1{,}429 \\
validation: synthetic & 1{,}000 & 1{,}385 \\
\bottomrule
\end{tabular}
\end{table}

\paragraph{Cleaning speech beds and noise.} The speech beds of the synthetic mixtures (\cref{sec:examples}) are taken from NVS-38K audio away from the labelled event and from NonverbalTTS clips without sounds, so they still contain unlabelled breaths, sniffs and laughter. Pasting events onto them would teach the model that such sounds are background. We therefore run a preliminary 31-class checkpoint of the same architecture (trained earlier on NVS-38K and Vaani only, with the same splits) over the beds and over the noise pool, mark frames where any speaker class (beds) or any class (noise) exceeds its tuned threshold, widen each mark by $\pm$0.32\,s, and keep the longest unmarked stretch of at least 2\,s. This kept 9{,}645 of 10{,}105 beds (18.4\,h) and 2{,}102 of 2{,}279 noise chunks (5.5\,h).

\section{Model}
\label{sec:model}

\begin{figure*}[!htbp]
\centering
\resizebox{\textwidth}{!}{\begin{tikzpicture}[font=\scriptsize, >={Latex[length=1.5mm]},
  op/.style={draw=sgray, rounded corners=2pt, align=center, inner sep=3pt, fill=white},
  arr/.style={->, sgray, line width=0.6pt}, lab/.style={font=\tiny, text=sgray, align=center}]
\node[op, text width=17mm, minimum height=13mm] (wav) at (0,1.1) {audio\\16\,kHz, 8\,s\\$128{,}000$ samples};
\node[lab, anchor=south] at (1.75,1.15) {STFT 512\\hop 10\,ms\\64 mel};
\draw[arr] (wav.east) -- (2.45,1.1);
\filldraw[fill=sline, draw=sgray, line width=0.35pt] (2.53,2.20) -- ++(0.14,0.08) -- ++(0.84,0) -- ++(-0.14,-0.08) -- cycle;
\filldraw[fill=sline!70!sgray, draw=sgray, line width=0.35pt] (3.37,0.00) -- ++(0.14,0.08) -- ++(0,2.20) -- ++(-0.14,-0.08) -- cycle;
\filldraw[fill=white, draw=sgray, line width=0.35pt] (2.53,0.00) rectangle ++(0.84,2.20);
\node[lab, anchor=north] at (3.02,-0.12) {\textbf{log-mel}\\$1\times64\times801$};
\filldraw[fill=sblue!11!white, draw=sgray, line width=0.35pt] (4.44,2.20) -- ++(0.38,0.22) -- ++(0.84,0) -- ++(-0.38,-0.22) -- cycle;
\filldraw[fill=sblue!21!black!25!white, draw=sgray, line width=0.35pt] (5.28,0.00) -- ++(0.38,0.22) -- ++(0,2.20) -- ++(-0.38,-0.22) -- cycle;
\filldraw[fill=sblue!21!white, draw=sgray, line width=0.35pt] (4.44,0.00) rectangle ++(0.84,2.20);
\node[lab, anchor=north] at (5.05,-0.12) {\textbf{block 1}\\$48\times64\times801$};
\draw[arr] (3.51,1.1) -- (4.42,1.1);
\node[lab, anchor=north] at (3.97,1.02) {};
\filldraw[fill=sblue!17!white, draw=sgray, line width=0.35pt] (6.58,1.73) -- ++(0.38,0.22) -- ++(0.84,0) -- ++(-0.38,-0.22) -- cycle;
\filldraw[fill=sblue!32!black!25!white, draw=sgray, line width=0.35pt] (7.42,0.48) -- ++(0.38,0.22) -- ++(0,1.25) -- ++(-0.38,-0.22) -- cycle;
\filldraw[fill=sblue!32!white, draw=sgray, line width=0.35pt] (6.58,0.48) rectangle ++(0.84,1.25);
\node[lab, anchor=north] at (7.19,-0.12) {\textbf{block 2}\\$48\times32\times801$};
\draw[arr] (5.65,1.1) -- (6.56,1.1);
\node[lab, anchor=north] at (6.12,1.02) {$f/2$};
\filldraw[fill=sblue!23!white, draw=sgray, line width=0.35pt] (8.73,1.49) -- ++(0.49,0.29) -- ++(0.84,0) -- ++(-0.49,-0.29) -- cycle;
\filldraw[fill=sblue!43!black!25!white, draw=sgray, line width=0.35pt] (9.57,0.71) -- ++(0.49,0.29) -- ++(0,0.77) -- ++(-0.49,-0.29) -- cycle;
\filldraw[fill=sblue!43!white, draw=sgray, line width=0.35pt] (8.73,0.71) rectangle ++(0.84,0.77);
\node[lab, anchor=north] at (9.39,-0.12) {\textbf{block 3}\\$96\times16\times801$};
\draw[arr] (7.80,1.1) -- (8.71,1.1);
\node[lab, anchor=north] at (8.26,1.02) {$f/2$};
\filldraw[fill=sblue!29!white, draw=sgray, line width=0.35pt] (10.99,1.37) -- ++(0.49,0.29) -- ++(0.53,0) -- ++(-0.49,-0.29) -- cycle;
\filldraw[fill=sblue!54!black!25!white, draw=sgray, line width=0.35pt] (11.52,0.83) -- ++(0.49,0.29) -- ++(0,0.54) -- ++(-0.49,-0.29) -- cycle;
\filldraw[fill=sblue!54!white, draw=sgray, line width=0.35pt] (10.99,0.83) rectangle ++(0.53,0.54);
\node[lab, anchor=north] at (11.50,-0.12) {\textbf{block 4}\\$96\times8\times401$};
\draw[arr] (10.06,1.1) -- (10.97,1.1);
\node[lab, anchor=north] at (10.52,1.02) {$f/2,\,t/2$};
\filldraw[fill=sblue!35!white, draw=sgray, line width=0.35pt] (12.94,1.31) -- ++(0.65,0.39) -- ++(0.38,0) -- ++(-0.65,-0.39) -- cycle;
\filldraw[fill=sblue!65!black!25!white, draw=sgray, line width=0.35pt] (13.31,0.89) -- ++(0.65,0.39) -- ++(0,0.42) -- ++(-0.65,-0.39) -- cycle;
\filldraw[fill=sblue!65!white, draw=sgray, line width=0.35pt] (12.94,0.89) rectangle ++(0.38,0.42);
\node[lab, anchor=north] at (13.45,-0.12) {\textbf{block 5}\\$192\times4\times201$};
\draw[arr] (12.01,1.1) -- (12.92,1.1);
\node[lab, anchor=north] at (12.47,1.02) {$f/2,\,t/2$};
\node[op, text width=17mm, minimum height=13mm, anchor=west] (flat) at (14.81,1.1) {flatten\\$192\times4=768$\\per 40\,ms step};
\draw[arr] (13.96,1.1) -- (flat.west);
\node[op, text width=21mm, minimum height=13mm] (proj) at (1.2,-2.0) {linear $768\rightarrow256$\\+ sinusoidal positions\\$201\times256$};
\draw[arr, rounded corners=3pt] (flat.south) -- ++(0,-1.35) -| (proj.north);
\node[op, fill=sorange!12, draw=sorange!65!black, text width=52mm, minimum height=13mm] (enc) at (6.0,-2.0) {\textbf{Transformer encoder}, 6 layers\\pre-norm; self-attention (8 heads)\\feed-forward $256{\to}1024{\to}256$, GELU\\$201\times256$ (Fig.~\ref{fig:pipeline}c)};
\node[op, fill=saqua!13, draw=saqua!65!black, text width=17mm, minimum height=13mm] (head) at (10.4,-2.0) {linear $256\rightarrow54$\\sigmoid};
\draw[arr] (proj) -- (enc); \draw[arr] (enc) -- (head);
\begin{scope}[shift={(11.95,-2.60)}]
\fill[sblue!5!white] (0.00,0.00) rectangle ++(0.085,0.19);
\fill[sblue!5!white] (0.09,0.00) rectangle ++(0.085,0.19);
\fill[sblue!8!white] (0.18,0.00) rectangle ++(0.085,0.19);
\fill[sblue!12!white] (0.27,0.00) rectangle ++(0.085,0.19);
\fill[sblue!3!white] (0.36,0.00) rectangle ++(0.085,0.19);
\fill[sblue!3!white] (0.45,0.00) rectangle ++(0.085,0.19);
\fill[sblue!12!white] (0.54,0.00) rectangle ++(0.085,0.19);
\fill[sblue!8!white] (0.63,0.00) rectangle ++(0.085,0.19);
\fill[sblue!5!white] (0.72,0.00) rectangle ++(0.085,0.19);
\fill[sblue!5!white] (0.81,0.00) rectangle ++(0.085,0.19);
\fill[sblue!12!white] (0.90,0.00) rectangle ++(0.085,0.19);
\fill[sblue!12!white] (0.99,0.00) rectangle ++(0.085,0.19);
\fill[sblue!12!white] (1.08,0.00) rectangle ++(0.085,0.19);
\fill[sblue!5!white] (1.17,0.00) rectangle ++(0.085,0.19);
\fill[sblue!5!white] (0.00,0.20) rectangle ++(0.085,0.19);
\fill[sblue!5!white] (0.09,0.20) rectangle ++(0.085,0.19);
\fill[sblue!12!white] (0.18,0.20) rectangle ++(0.085,0.19);
\fill[sblue!3!white] (0.27,0.20) rectangle ++(0.085,0.19);
\fill[sblue!85!white] (0.36,0.20) rectangle ++(0.085,0.19);
\fill[sblue!5!white] (0.45,0.20) rectangle ++(0.085,0.19);
\fill[sblue!8!white] (0.54,0.20) rectangle ++(0.085,0.19);
\fill[sblue!60!white] (0.63,0.20) rectangle ++(0.085,0.19);
\fill[sblue!5!white] (0.72,0.20) rectangle ++(0.085,0.19);
\fill[sblue!12!white] (0.81,0.20) rectangle ++(0.085,0.19);
\fill[sblue!3!white] (0.90,0.20) rectangle ++(0.085,0.19);
\fill[sblue!8!white] (0.99,0.20) rectangle ++(0.085,0.19);
\fill[sblue!12!white] (1.08,0.20) rectangle ++(0.085,0.19);
\fill[sblue!12!white] (1.17,0.20) rectangle ++(0.085,0.19);
\fill[sblue!12!white] (0.00,0.40) rectangle ++(0.085,0.19);
\fill[sblue!12!white] (0.09,0.40) rectangle ++(0.085,0.19);
\fill[sblue!12!white] (0.18,0.40) rectangle ++(0.085,0.19);
\fill[sblue!5!white] (0.27,0.40) rectangle ++(0.085,0.19);
\fill[sblue!8!white] (0.36,0.40) rectangle ++(0.085,0.19);
\fill[sblue!3!white] (0.45,0.40) rectangle ++(0.085,0.19);
\fill[sblue!3!white] (0.54,0.40) rectangle ++(0.085,0.19);
\fill[sblue!5!white] (0.63,0.40) rectangle ++(0.085,0.19);
\fill[sblue!12!white] (0.72,0.40) rectangle ++(0.085,0.19);
\fill[sblue!5!white] (0.81,0.40) rectangle ++(0.085,0.19);
\fill[sblue!8!white] (0.90,0.40) rectangle ++(0.085,0.19);
\fill[sblue!12!white] (0.99,0.40) rectangle ++(0.085,0.19);
\fill[sblue!8!white] (1.08,0.40) rectangle ++(0.085,0.19);
\fill[sblue!12!white] (1.17,0.40) rectangle ++(0.085,0.19);
\fill[sblue!12!white] (0.00,0.60) rectangle ++(0.085,0.19);
\fill[sblue!8!white] (0.09,0.60) rectangle ++(0.085,0.19);
\fill[sblue!12!white] (0.18,0.60) rectangle ++(0.085,0.19);
\fill[sblue!5!white] (0.27,0.60) rectangle ++(0.085,0.19);
\fill[sblue!8!white] (0.36,0.60) rectangle ++(0.085,0.19);
\fill[sblue!3!white] (0.45,0.60) rectangle ++(0.085,0.19);
\fill[sblue!8!white] (0.54,0.60) rectangle ++(0.085,0.19);
\fill[sblue!5!white] (0.63,0.60) rectangle ++(0.085,0.19);
\fill[sblue!8!white] (0.72,0.60) rectangle ++(0.085,0.19);
\fill[sblue!3!white] (0.81,0.60) rectangle ++(0.085,0.19);
\fill[sblue!5!white] (0.90,0.60) rectangle ++(0.085,0.19);
\fill[sblue!8!white] (0.99,0.60) rectangle ++(0.085,0.19);
\fill[sblue!8!white] (1.08,0.60) rectangle ++(0.085,0.19);
\fill[sblue!3!white] (1.17,0.60) rectangle ++(0.085,0.19);
\fill[sblue!3!white] (0.00,0.80) rectangle ++(0.085,0.19);
\fill[sblue!12!white] (0.09,0.80) rectangle ++(0.085,0.19);
\fill[sblue!12!white] (0.18,0.80) rectangle ++(0.085,0.19);
\fill[sblue!3!white] (0.27,0.80) rectangle ++(0.085,0.19);
\fill[sblue!12!white] (0.36,0.80) rectangle ++(0.085,0.19);
\fill[sblue!5!white] (0.45,0.80) rectangle ++(0.085,0.19);
\fill[sblue!20!white] (0.54,0.80) rectangle ++(0.085,0.19);
\fill[sblue!8!white] (0.63,0.80) rectangle ++(0.085,0.19);
\fill[sblue!5!white] (0.72,0.80) rectangle ++(0.085,0.19);
\fill[sblue!12!white] (0.81,0.80) rectangle ++(0.085,0.19);
\fill[sblue!12!white] (0.90,0.80) rectangle ++(0.085,0.19);
\fill[sblue!12!white] (0.99,0.80) rectangle ++(0.085,0.19);
\fill[sblue!3!white] (1.08,0.80) rectangle ++(0.085,0.19);
\fill[sblue!3!white] (1.17,0.80) rectangle ++(0.085,0.19);
\fill[sblue!3!white] (0.00,1.00) rectangle ++(0.085,0.19);
\fill[sblue!12!white] (0.09,1.00) rectangle ++(0.085,0.19);
\fill[sblue!8!white] (0.18,1.00) rectangle ++(0.085,0.19);
\fill[sblue!8!white] (0.27,1.00) rectangle ++(0.085,0.19);
\fill[sblue!5!white] (0.36,1.00) rectangle ++(0.085,0.19);
\fill[sblue!3!white] (0.45,1.00) rectangle ++(0.085,0.19);
\fill[sblue!8!white] (0.54,1.00) rectangle ++(0.085,0.19);
\fill[sblue!3!white] (0.63,1.00) rectangle ++(0.085,0.19);
\fill[sblue!3!white] (0.72,1.00) rectangle ++(0.085,0.19);
\fill[sblue!3!white] (0.81,1.00) rectangle ++(0.085,0.19);
\fill[sblue!3!white] (0.90,1.00) rectangle ++(0.085,0.19);
\fill[sblue!5!white] (0.99,1.00) rectangle ++(0.085,0.19);
\fill[sblue!12!white] (1.08,1.00) rectangle ++(0.085,0.19);
\fill[sblue!8!white] (1.17,1.00) rectangle ++(0.085,0.19);
\draw[sgray, line width=0.3pt] (0,0) rectangle (1.26,1.2);
\node[lab, anchor=west] at (1.35,0.6) {probabilities\\$201\times54$\\every 40\,ms};
\end{scope}
\draw[arr] (head.east) -- (11.9,-2.0);
\end{tikzpicture}}
\caption{Architecture, with tensor shapes measured on the released model. \textbf{Top:} the waveform becomes a 64-band log-mel spectrogram, which five convolutional blocks turn into 201 time steps of 768 features. The blocks are drawn to scale (height: frequency bins; width: time steps; depth: square root of the channels), with shapes channels\,$\times$\,frequency\,$\times$\,time below each and strides under the arrows; \cref{fig:pipeline}c shows the content of a block. \textbf{Bottom:} the features are projected to 256 dimensions, given sinusoidal positions and passed through six pre-norm Transformer encoder layers; a linear layer with a sigmoid gives one probability per class every 40\,ms.}
\label{fig:arch}
\end{figure*}
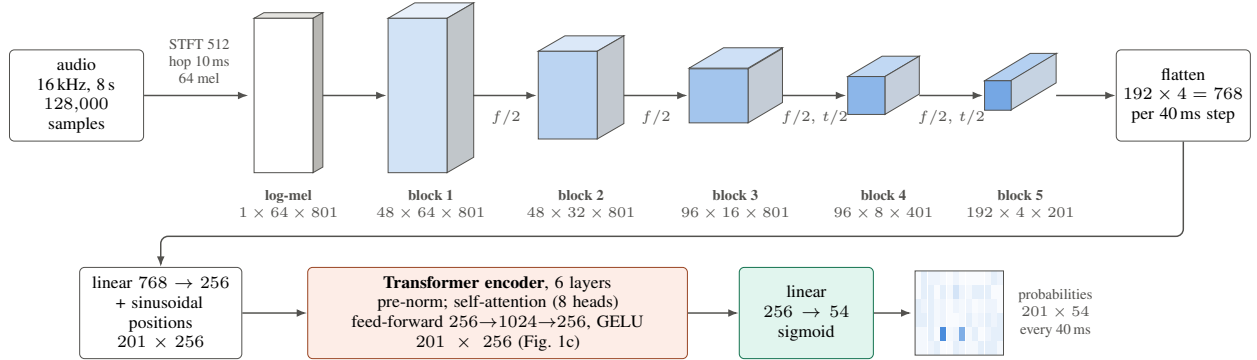

\Cref{fig:arch} shows the network and \cref{tab:layers} lists every layer with its output shape and parameter count; shapes and counts were measured with forward hooks on the released model.

\paragraph{Front end.} An 8\,s window (128{,}000 samples at 16\,kHz) is converted to a 64-band log-mel spectrogram (512-point FFT, 25\,ms window, 10\,ms hop, 50--8000\,Hz), compressed as $10\log_{10}(\cdot+10^{-9})$ and scaled to $(x+60)/20$ clipped to $[-1,3]$; this yields $64\times801$ frames.

\paragraph{Convolutional blocks.} Five blocks, each two $3\times3$ convolutions with batch normalization~\citep{ioffe2015batchnorm} and ReLU, have 48, 48, 96, 96 and 192 channels. The first convolution of blocks 2 and 3 has stride 2 along frequency, and that of blocks 4 and 5 stride 2 along both axes, so frequency shrinks from 64 bins to 4 and time from 801 frames of 10\,ms to 201 steps of 40\,ms. Each output step of the front end depends on 31 mel frames, i.e.\ about 0.33\,s of audio.

\paragraph{Transformer and head.} The $192\times4$ features of each step are flattened to 768, projected to 256 dimensions, summed with fixed sinusoidal position encodings~\citep{vaswani2017attention}, and passed through six pre-norm~\citep{xiong2020layer} Transformer encoder layers, each with eight-head self-attention and a feed-forward block of 1{,}024 units with GELU~\citep{hendrycks2016gelu} and dropout 0.1. Every step can therefore draw on the whole 8\,s window. A linear layer with a sigmoid per class gives 201 output frames of 40\,ms for each of the 54 classes.

\begin{table*}[!htbp]
\centering\small
\caption{Layer-by-layer view of the network for one 8\,s window (batch dimension omitted). Shapes are channels $\times$ frequency $\times$ time for the convolutional part and time $\times$ features afterwards; they were measured with forward hooks on the released model. Strides are (frequency, time). Parameter counts are exact.}
\label{tab:layers}
\begin{tabularx}{\textwidth}{@{}lXrr@{}}
\toprule
Layer & Operation & Output shape & Parameters \\
\midrule
input & waveform, 16 kHz, 8 s & $128000$ & -- \\
log-mel & STFT 512 / win 400 / hop 160, 64 mel bands, log, scale & $64\times801$ & -- \\
conv block 1 & 2 $\times$ (conv 3$\times$3, BN, ReLU); 1$\rightarrow$48 ch; stride (f,t)=(1, 1) & $48\times64\times801$ & 21{,}456 \\
conv block 2 & 2 $\times$ (conv 3$\times$3, BN, ReLU); 48$\rightarrow$48 ch; stride (f,t)=(2, 1) & $48\times32\times801$ & 41{,}760 \\
conv block 3 & 2 $\times$ (conv 3$\times$3, BN, ReLU); 48$\rightarrow$96 ch; stride (f,t)=(2, 1) & $96\times16\times801$ & 124{,}992 \\
conv block 4 & 2 $\times$ (conv 3$\times$3, BN, ReLU); 96$\rightarrow$96 ch; stride (f,t)=(2, 2) & $96\times8\times401$ & 166{,}464 \\
conv block 5 & 2 $\times$ (conv 3$\times$3, BN, ReLU); 96$\rightarrow$192 ch; stride (f,t)=(2, 2) & $192\times4\times201$ & 498{,}816 \\
reshape & (C $\times$ F) per time step: 192 $\times$ 4 = 768 & $201\times768$ & -- \\
projection & linear 768 $\rightarrow$ 256, + sinusoidal positions & $201\times256$ & 196{,}864 \\
encoder layer 1 & pre-norm MHSA (8 heads) + FF 1024, GELU & $201\times256$ & 789{,}760 \\
encoder layer 2 & pre-norm MHSA (8 heads) + FF 1024, GELU & $201\times256$ & 789{,}760 \\
encoder layer 3 & pre-norm MHSA (8 heads) + FF 1024, GELU & $201\times256$ & 789{,}760 \\
encoder layer 4 & pre-norm MHSA (8 heads) + FF 1024, GELU & $201\times256$ & 789{,}760 \\
encoder layer 5 & pre-norm MHSA (8 heads) + FF 1024, GELU & $201\times256$ & 789{,}760 \\
encoder layer 6 & pre-norm MHSA (8 heads) + FF 1024, GELU & $201\times256$ & 789{,}760 \\
head & linear 256 $\rightarrow$ 54, sigmoid & $201\times54$ & 13{,}878 \\
\midrule
\multicolumn{3}{@{}l}{total: convolutional front end 853{,}488, projection 196{,}864, Transformer 4{,}738{,}560, head 13{,}878} & 5{,}802{,}790 \\
\bottomrule
\end{tabularx}
\end{table*}

\begin{figure}[!htbp]
\centering
\includegraphics[width=0.6\linewidth]{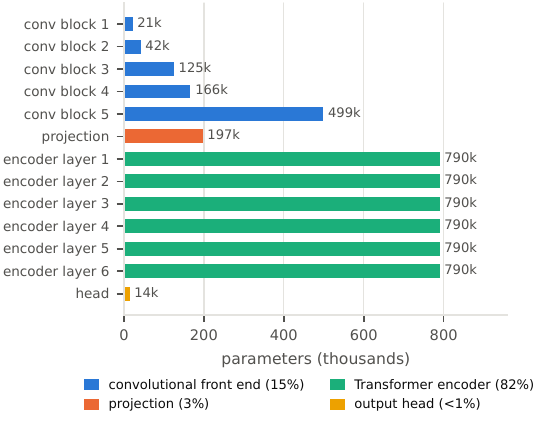}
\caption{Parameters per layer, coloured by component (5{,}802{,}790 in total).}
\label{fig:params}
\end{figure}

\paragraph{Size.} The model has 5{,}802{,}790 trainable parameters (27\,MB as \texttt{safetensors}). \Cref{fig:params} shows where they are: 82\,\% in the six encoder layers (789{,}760 each: 263{,}168 in self-attention, 525{,}568 in the feed-forward block and 1{,}024 in the two layer norms), 15\,\% in the convolutions, 3\,\% in the projection and under 1\,\% in the output layer. The design is deliberately small: it trains in about three hours on one T4 and runs far faster than real time on a CPU (\cref{sec:speed}).

\begin{figure*}[!htbp]
\centering
\includegraphics[width=\textwidth]{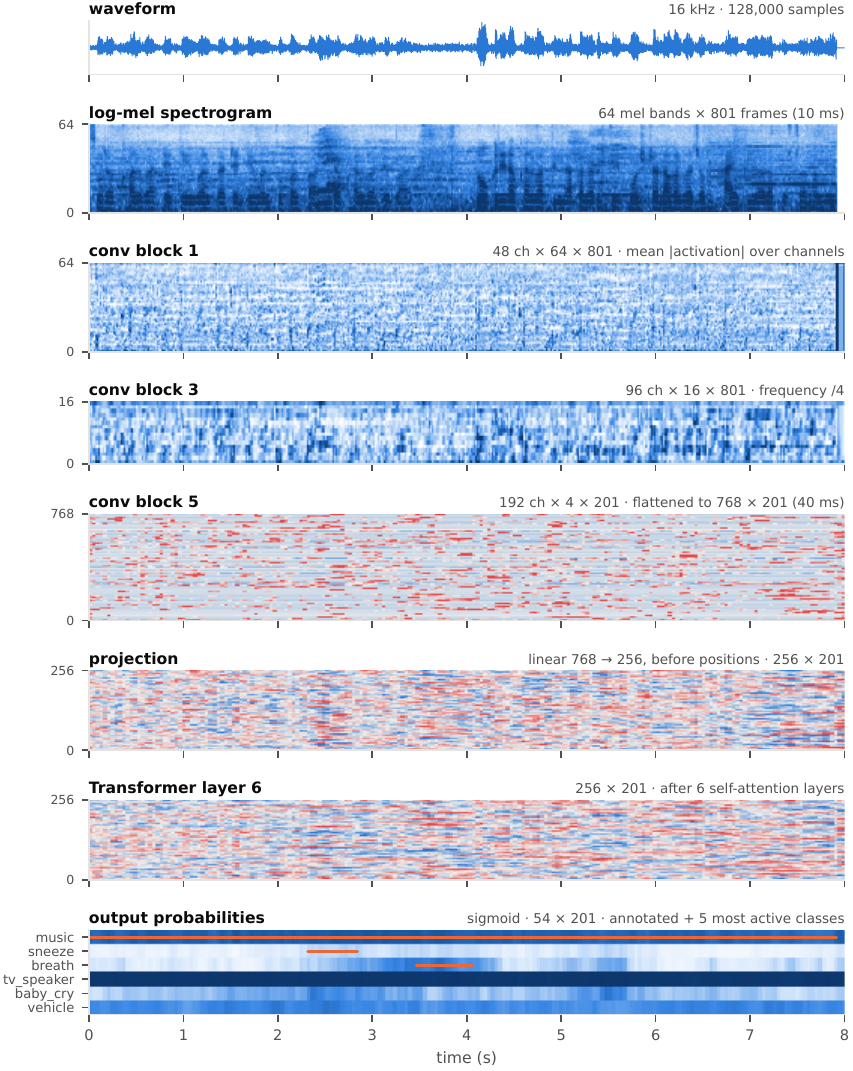}
\caption{A held-out clip passing through the released model; all panels share the time axis. Convolutional panels show the mean absolute activation over channels (blue scale); vector-valued panels show each feature standardized over time (blue below, red above its mean). The bottom panel shows the output probabilities of the annotated classes and the five most active classes (white 0, dark blue 1), with the annotated events as orange lines. The dark column at the right edge of block 1 is the zero padding of this 7.9\,s clip to the 8\,s window.}
\label{fig:signalflow}
\end{figure*}

\paragraph{How a recording passes through the network.} \Cref{fig:signalflow} follows one held-out clip (the clip of \cref{fig:example}) through the released model, showing the real intermediate representations. The spectrogram's fine frequency structure is still visible after the first block; by block 3 the frequency axis has four times fewer bins; after block 5 each 40\,ms step is a 768-dimensional vector, which the projection and the six Transformer layers transform into 256-dimensional context-dependent representations. The last panel shows the resulting class probabilities, with the annotated events overlaid.

\section{Training}
\label{sec:training}

\subsection{Training examples}
\label{sec:examples}
All training examples are 8\,s windows with frame-exact labels. Two generators produce them.

\paragraph{Synthetic mixtures.} A cleaned speech bed (\cref{sec:banks}) is tiled or cropped to 8\,s and scaled to a reference RMS of 0.05; with probability 0.10 no bed is used. With probability 0.85 the window receives one, two or three events (with probabilities $\tfrac12$, $\tfrac13$, $\tfrac16$); each event's class is drawn uniformly over classes, so rare classes are as frequent as common ones, and a crop of that class is drawn from all sources. Crops longer than 8\,s are randomly cut. With probability 0.3 a crop is speed-perturbed by a factor in $[0.9,1.1]$. It is placed at a uniformly random time, at a speech-to-event ratio drawn from $[-3,12]$\,dB, or 10--25\,dB below the reference level when there is no speech. The whole mixture is scaled by $[-10,+6]$\,dB and a small noise floor is added. Because the generator places the events, their positions are known by construction~\citep{salamon2017scaper,turpault2019desed}, up to the accuracy of the crop boundaries in the source annotations.

\paragraph{Real windows.} Real annotated clips of one source (NVS-38K, Vaani, NonverbalTTS or AudioSet-Strong) are concatenated, starting at a random offset in the first, until 8\,s are filled. Each clip is normalized to an RMS level that varies by $\pm6$\,dB, joins get 10\,ms fades, and the annotated events are shifted accordingly, keeping event parts of at least 50\,ms. Real windows teach the model what the events sound like in their natural context, which synthetic pasting cannot fully provide.

\paragraph{Augmentation.} Every example is stored once clean and, in addition, once (synthetic) or twice (real) in an independently augmented version (\cref{tab:augment}). Inserted silences shift later labels. Background noise comes 40\,\% of the time from the real noise pool (DNC, DEMAND), and otherwise from babble made of 3--6 speech beds, synthetic mains hum with harmonics, or white, pink or brown noise. The phone-band filter and $\mu$-law codec~\citep{itu1988g711} mimic telephone audio.

\begin{table}[!htbp]
\centering\small
\caption{Augmentations applied to each stored augmented copy; each is drawn independently.}
\label{tab:augment}
\begin{tabularx}{\linewidth}{@{}>{\raggedright\arraybackslash}p{0.33\linewidth}c>{\raggedright\arraybackslash}X@{}}
\toprule
Augmentation & Prob. & Parameters \\
\midrule
silence inserted outside events, labels shifted & 0.5 & 1--2 pauses of 0.2--1.5\,s \\
reverberation (exponentially decaying noise response) & 0.25 & RT$_{60}$ 0.15--0.8\,s; direct path 1--4$\times$ the tail peak \\
telephone band & 0.2 & Butterworth band-pass 300--3400\,Hz (order 4 per edge); 8-bit $\mu$-law with probability 0.5 \\
background noise & 0.85 & real noise 40\,\% (SNR 3--25\,dB), babble 20\,\% (12--30\,dB), hum 10\,\% (10--35\,dB), white/pink/brown 30\,\% (8--35\,dB) \\
gain & 1 & $-12$ to $+6$\,dB \\
\bottomrule
\end{tabularx}
\end{table}

\paragraph{Pools and features.} Fixed pools are generated once and stored as fp16 log-mel features: 30{,}000 synthetic mixtures and 9{,}000 (NVS-38K), 9{,}000 (Vaani), 7{,}000 (NonverbalTTS) and 12{,}000 (AudioSet-Strong) real windows, which with the augmented copies gives \textbf{171{,}000 training examples} (\cref{tab:banks}). Training reads only the cached features, so the CPU does not limit throughput. During training, each batch is circularly shifted in time by a random number of output frames (labels shifted with it), and with probability 0.5 per batch SpecAugment~\citep{park2019specaugment} masks, in each example, one time band of 1--29 mel frames (10--290\,ms) and one frequency band of 1--7 mel bins.

\subsection{Objective and optimization}
\label{sec:objective}
Frame targets $y_{t,c}\in\{0,1\}$ are derived from the event lists and dilated by one frame (40\,ms) on each side. With logits $z_{t,c}$, the loss is
\begin{equation}
\mathcal{L} = \frac{1}{TC}\sum_{t,c} \mathrm{BCE}_{w_c}\!\left(z_{t,c}, y_{t,c}\right) \;+\; 0.25\,\frac{1}{C}\sum_c \mathrm{BCE}\!\left(\max_t z_{t,c},\, \max_t y_{t,c}\right),
\label{eq:loss}
\end{equation}
where $\mathrm{BCE}_{w}$ weights the positive term by $w_c = \mathrm{clip}\big(N^{-}_c/N^{+}_c,\,1,\,100\big)$, the ratio of negative to positive frames of class $c$ in the training pools. The second term is a clip-level loss on the maximum over time.

We train for 10 epochs with AdamW~\citep{loshchilov2019adamw} (weight decay 0.01), a OneCycle schedule~\citep{smith2019superconvergence} with peak learning rate $3\times10^{-4}$ and 10\,\% warm-up, batch size 32, mixed precision and gradient-norm clipping at 5 (\cref{tab:config}). The network is warm-started from the preliminary 31-class checkpoint of \cref{sec:banks}: all weights except the output layer are copied, as are the output rows of its 31 classes; the other 23 classes start from random initialization. That checkpoint was trained only on NVS-38K and Vaani with the same held-out splits, so it never saw any evaluation clip. One epoch of 5{,}343 steps takes about 1{,}095\,s on a T4.

\begin{table}[!htbp]
\centering\small
\caption{Training configuration of the released model.}
\label{tab:config}
\begin{tabular}{@{}ll@{}}
\toprule
Setting & Value \\
\midrule
Input & 16\,kHz mono, 8\,s windows; 64-band log-mel, 25\,ms window, 10\,ms hop, 50--8000\,Hz \\
Output & 54 sigmoid outputs every 40\,ms (201 frames per window) \\
Parameters & 5{,}802{,}790 trainable (plus a fixed sinusoidal position table) \\
Optimizer & AdamW, weight decay 0.01, gradient-norm clip 5 \\
Schedule & OneCycle, peak LR 0.0003, 10\,\% warm-up, 10 epochs \\
Batch & 32 windows; 5{,}343 steps per epoch \\
Precision & mixed (fp16 autocast), features cached as fp16 \\
Initialisation & warm start from a preliminary 31-class checkpoint (all weights except the output layer; its 31 output rows) \\
Per-class crop cap & about 3{,}000 per source (train), 1/8 for validation; NonverbalTTS uncapped \\
Real-window caps & NVS-38K 15{,}000, Vaani 20{,}000, AudioSet 30{,}000 \\
AudioSet streaming & up to 500 training + 60 evaluation row groups \\
Pools & 30{,}000 synthetic; real NVS-38K 9{,}000, Vaani 9{,}000, NonverbalTTS 7{,}000, AudioSet 12{,}000 \\
Hardware & one NVIDIA T4 (16\,GB), 16 vCPU; $\approx$1{,}095\,s per epoch \\
\bottomrule
\end{tabular}
\end{table}

\paragraph{Model selection.} After every epoch, frame probabilities are computed for fixed validation windows: 600 NVS-38K, 800 Vaani, 400 NonverbalTTS and 800 AudioSet-Strong real windows, 600 augmented (``noisy'') real windows, and 1{,}000 synthetic mixtures. For each set we compute the macro average, over the classes that set's source labels (all classes for the noisy set, which mixes the four sources) that have at least one positive frame, of the frame-level average precision
\begin{equation}
\mathrm{AP}_c = \frac{1}{N^{+}_c}\sum_{k=1}^{N} \mathrm{Prec}_c(k)\,\mathbb{1}[y_{(k),c}=1],
\label{eq:ap}
\end{equation}
where frames are sorted by decreasing $p_{t,c}$, $y_{(k),c}$ is the label of the $k$-th frame and $\mathrm{Prec}_c(k)$ the precision of the top $k$. Validation targets are dilated by one frame like the training targets. The checkpoint with the highest mean over the five real or noisy sets is kept; synthetic data does not influence selection. AP does not depend on a threshold, so selection rewards a model that ranks real events above everything else, and the operating point is chosen afterwards.

\subsection{Calibration: two threshold profiles}
\label{sec:calibration}
For each class, the frame probabilities of the real validation windows of the sources that label it are pooled. A threshold $\theta_c$ is chosen from $\{0.05, 0.075, \dots, 0.975\}$ to maximize frame F1 (the \emph{balanced} profile) or frame F2, which weights recall twice as much as precision (the \emph{sensitive} profile). To estimate the performance of a chosen threshold without bias from the choice, we split the windows into even and odd halves, tune on one half, score on the other, swap, and average: this two-fold cross-validated (CV) F1 or F2 is what we report. The released thresholds are tuned on all windows. A class with fewer than 25 positive frames (1\,s) or a CV F1 below 0.1 is marked \emph{weak}; its balanced threshold is raised to at least 0.9 and its sensitive threshold clamped to $[0.6, 0.8]$, so that weak classes stay available but fire only when the model is confident.

\section{Inference and transcript merge}
\label{sec:inference}

\subsection{From probabilities to events}
\label{sec:postproc}
Audio of any length is cut into 8\,s windows with a 2\,s hop (the last window zero-padded), and the probabilities of overlapping windows are averaged per 40\,ms frame. Each class track is median-filtered over five frames (200\,ms). An event is a maximal run of frames above $0.6\,\theta_c$ that contains at least one frame above $\theta_c$ (hysteresis), so that brief dips do not split one event into several. Runs of the same class closer than 0.3\,s (speaker sounds) or 1.0\,s (background sounds) are joined, and events shorter than 0.15\,s are discarded. A threshold above 1 disables a class.

\subsection{Merging events into a transcript}
\label{sec:merge}
The merge takes the recognizer's words with timestamps, $(w_i, s_i, e_i)$, and the detected events.
\begin{enumerate}
\item Events the recognizer already transcribed are removed: an event is dropped if every word overlapping it is a filler (\emph{uh, um, hmm, \dots}) or a laughter token (\emph{ha, haha, \dots}).
\item Same-class pieces closer than 0.5\,s (speaker) or 1.5\,s (background) are joined.
\item Word start times are made non-decreasing ($s'_i = \max_{j\le i} s_j$). A speaker sound or the opening tag of a background sound is inserted before the first word with $s'_i \ge t_\text{start}$; a closing tag is inserted after the last word with $s'_i \le t_\text{end}$, and only if the event overlaps at least one word and ends at least 0.3\,s before the end of the last word.
\item Identical adjacent tags are collapsed.
\end{enumerate}
Because tags are inserted between words rather than sorted together with them, the recognizer's word order is never changed, even when its timestamps are not monotonic. The result is shown in the example of \cref{sec:task}.

\section{Experiments}
\label{sec:eval}

\subsection{Protocol}
\label{sec:protocol}
All evaluation uses real recordings held out from training (\cref{sec:splits}), and each class is scored only on the sources that annotate it. NVS-38K, for example, has no labels for horns, so a horn predicted in an NVS-38K clip is neither a hit nor a false alarm. We report three levels:
\begin{enumerate}
\item \textbf{Frame-level average precision} on the fixed validation windows (\cref{eq:ap}). It is threshold-free and is also used for model selection.
\item \textbf{Cross-validated frame F1/F2, precision and recall} at the tuned thresholds (\cref{sec:calibration}).
\item \textbf{Event-level recall, precision and F1} on whole held-out clips (up to 1{,}500 per source) run through the complete detector of \cref{sec:postproc}. A true event counts as \emph{found} if a prediction of the same class overlaps it in time, and a prediction counts as \emph{correct} if it overlaps a true event of the same class. This criterion ignores boundary accuracy, which the frame-level measures capture; it is more lenient than collar-based event metrics~\citep{mesaros2016metrics}.
\end{enumerate}
\paragraph{Caveats.} (i) Checkpoint selection, the threshold tuning and the frame-level measures all use validation windows drawn from the same held-out clips as the event-level evaluation, so absolute numbers are optimistic. (ii) No standard benchmark, such as the DCASE domestic SED task~\citep{turpault2019desed}, was run. (iii) All numbers come from a single training run without repetitions, so we cannot give confidence intervals. (iv) Sounds that annotators did not label count as false alarms, which penalizes classes such as \evt{breath} whose instances are often unlabelled.

\subsection{Training dynamics}
\begin{figure}[!htbp]
\centering
\includegraphics[width=\linewidth]{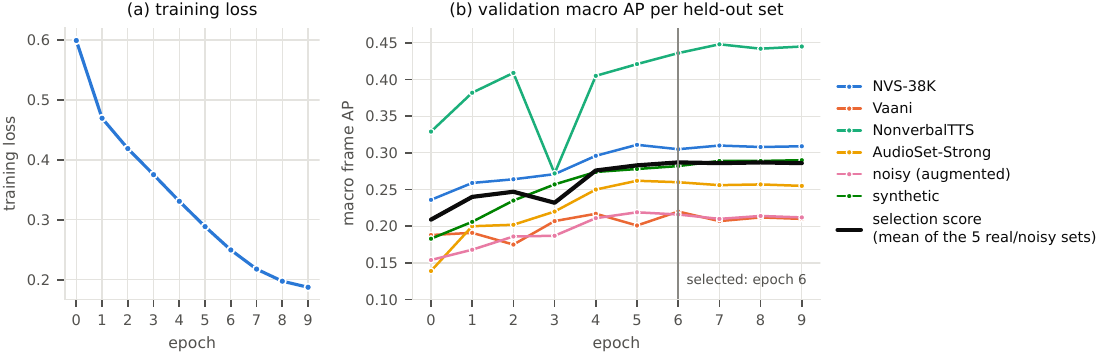}
\caption{Training loss (a) and validation macro frame AP per held-out set (b). The selection score is the mean over the four real sets and the noisy set; the synthetic set is shown but not used for selection. Values are in \texttt{data/training\_log.csv}.}
\label{fig:training}
\end{figure}

\Cref{fig:training} shows the training loss and the validation macro AP per held-out set. Epochs are numbered from 0. The selection score rises from 0.209 after epoch 0 to 0.287 after epoch 6 and stays between 0.286 and 0.287 afterwards, while the training loss keeps falling from 0.250 to 0.188. The model therefore stops improving on real held-out audio after about seven epochs, and further epochs mainly fit the fixed training pools. AudioSet-Strong improves the most during training (0.139 to 0.260 at the selected epoch), Vaani the least (0.188 to 0.220). The NonverbalTTS set dips at epoch 3 (0.409 to 0.272) and recovers at the next epoch; with 400 windows, a single epoch's value on this set is noisy. The checkpoint from epoch 6 (the seventh epoch) is the released model.

\subsection{Per-class frame-level results}
\Cref{fig:classap} shows the frame AP and the cross-validated F1 of every class; \cref{tab:perclass} in the appendix lists all values. The macro AP is 0.265 over all 54 classes (\evt{knock}, with no positive validation frames, counts as 0) and 0.324 over the 43 classes that are not weak. The highest APs are for \evt{music} (0.884), \evt{sigh} (0.725), \evt{horn} (0.590), \evt{applause} (0.582) and \evt{snore} (0.551). Eleven classes are weak: \evt{tv\_speaker}, \evt{fan}, \evt{footsteps}, \evt{cry}, \evt{groan}, \evt{chew}, \evt{door}, \evt{yawn}, \evt{hiccup}, \evt{nose\_blow} and \evt{knock}. Some of them have hardly any positive validation frames (\evt{knock} has none, \evt{nose\_blow} 36); others, such as \evt{tv\_speaker} and \evt{fan}, have more validation data but still score low; we have not analysed their confusions.

\begin{figure}[!htbp]
\centering
\includegraphics[width=\linewidth]{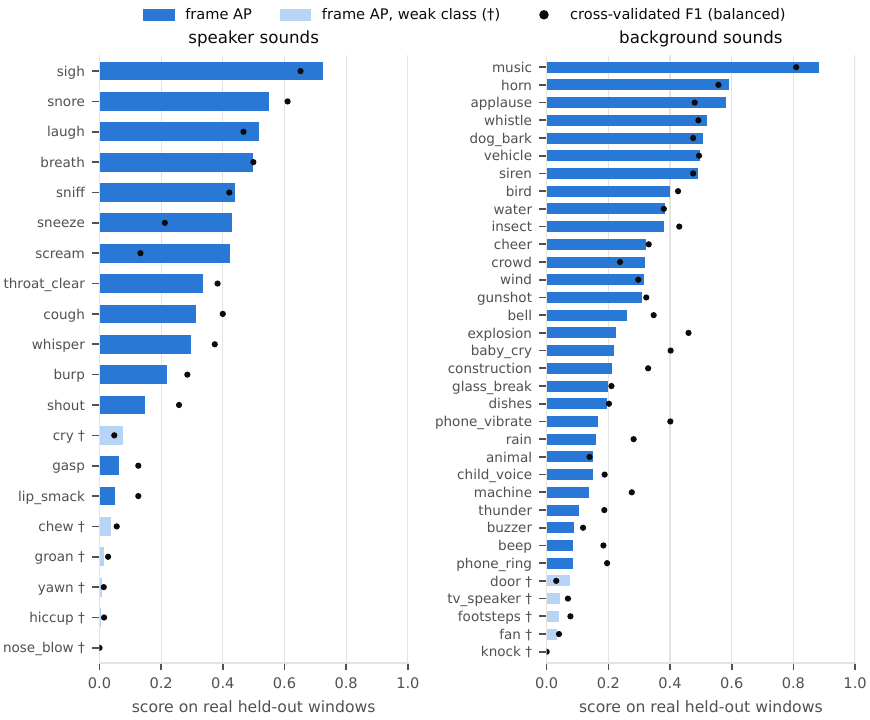}
\caption{Frame-level average precision (bars) and two-fold cross-validated F1 of the balanced profile (dots; thresholds tuned on the other half of the windows) per class, on real held-out windows. $\dagger$ marks weak classes (\cref{sec:calibration}).}
\label{fig:classap}
\end{figure}

\subsection{Operating profiles}
\begin{figure}[!htbp]
\centering
\includegraphics[width=0.55\linewidth]{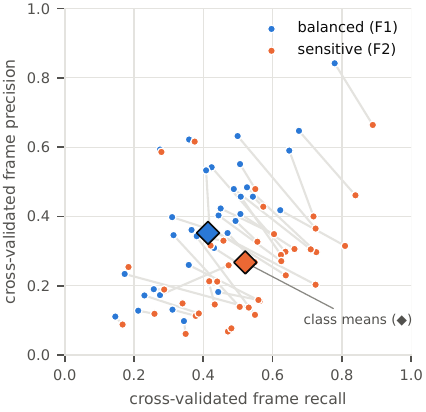}
\caption{Cross-validated frame precision and recall of each reliable class under the balanced (F1) and sensitive (F2) thresholds; grey lines connect the two points of a class. Diamonds are the class means.}
\label{fig:profiles}
\end{figure}

\Cref{tab:profiles} and \cref{fig:profiles} compare the two profiles on the 43 reliable classes. Moving from the balanced to the sensitive profile raises mean cross-validated recall from 0.414 to 0.521 and lowers mean precision from 0.352 to 0.267. Users who review tags, or who prefer extra tags to missed sounds, can use the sensitive profile; the balanced profile suits unattended transcription.

\begin{table}[!htbp]
\centering\small
\caption{Cross-validated frame-level operating points of the two threshold profiles, averaged over the 43 reliably measured classes.}
\label{tab:profiles}
\begin{tabular}{@{}lccc@{}}
\toprule
Profile & Precision & Recall & Objective \\
\midrule
balanced & 0.352 & 0.414 & F1 = 0.351 \\
sensitive & 0.267 & 0.521 & F2 = 0.400 \\
\bottomrule
\end{tabular}
\end{table}

\subsection{Event-level results}
\begin{figure}[!htbp]
\centering
\includegraphics[width=\linewidth]{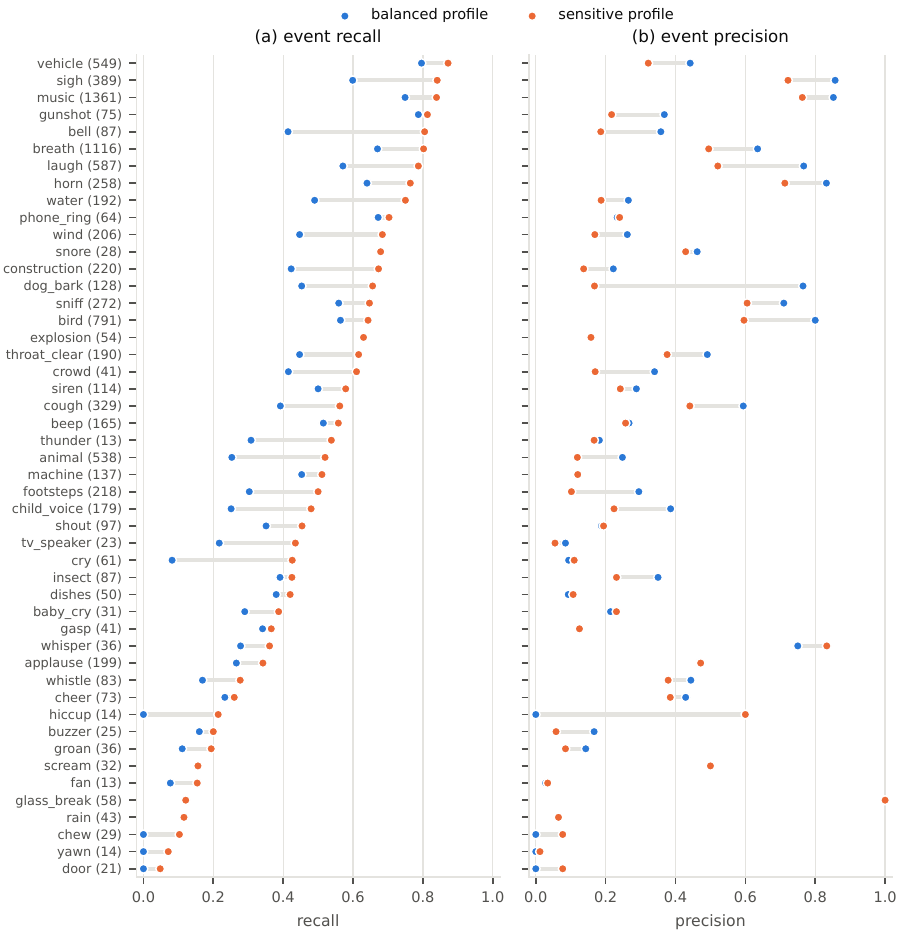}
\caption{Event-level recall (a) and precision (b) per class on whole held-out clips with the balanced and sensitive profiles; held-out event counts in parentheses. Classes with at least 10 events, sorted by sensitive recall. Values are in \texttt{data/event\_eval.csv}.}
\label{fig:eventeval}
\end{figure}

\Cref{tab:main} summarizes the event-level results on whole held-out clips and \cref{fig:eventeval} shows them per class. Over the 48 classes with at least 10 held-out events (9{,}367 events), the sensitive profile reaches a macro recall of 0.491 at a macro precision of 0.302 (F1 0.322) and finds 67\,\% of all events; the balanced profile reaches 0.369 recall at 0.361 precision (F1 0.329). Speaker sounds are found about as often as background sounds (macro recall 0.455 vs.\ 0.510 with the sensitive profile) but with higher precision (0.383 vs.\ 0.261). The best classes by event F1 (sensitive) are \evt{music} (0.80), \evt{sigh} (0.78), \evt{horn} (0.74), \evt{laugh} (0.63), \evt{sniff} (0.63), \evt{bird} (0.62) and \evt{breath} (0.61); seven classes are below 0.1: \evt{yawn}, \evt{fan}, \evt{door}, \evt{rain}, \evt{chew}, \evt{buzzer} and \evt{tv\_speaker}. For 26 of the 48 classes, the sensitive profile finds at least half of the events.

\begin{table*}[!htbp]
\centering\small
\caption{Event-level results on whole held-out real clips, for the sensitive (S) and balanced (B) threshold profiles. Macro averages over classes with at least 10 held-out events; the last two columns give the event-weighted (micro) recall. A true event is found if a same-class prediction overlaps it; a prediction is correct if it overlaps a same-class true event. Thresholds were tuned on held-out data that overlaps these clips, so absolute values are optimistic (\cref{sec:protocol}).}
\label{tab:main}
\begin{tabular}{@{}lrr cc cc cc cc@{}}
\toprule
 & & & \multicolumn{2}{c}{Recall} & \multicolumn{2}{c}{Precision} & \multicolumn{2}{c}{F1} & \multicolumn{2}{c}{Micro recall} \\
\cmidrule(lr){4-5}\cmidrule(lr){6-7}\cmidrule(lr){8-9}\cmidrule(l){10-11}
Classes & $n$ & Events & S & B & S & B & S & B & S & B \\
\midrule
speaker sounds & 16 & 3{,}271 & 0.455 & 0.327 & 0.383 & 0.395 & 0.378 & 0.338 & 0.703 & 0.542 \\
background sounds & 32 & 6{,}096 & 0.510 & 0.390 & 0.261 & 0.345 & 0.294 & 0.325 & 0.656 & 0.528 \\
\midrule
all & 48 & 9{,}367 & 0.491 & 0.369 & 0.302 & 0.361 & 0.322 & 0.329 & 0.672 & 0.533 \\
\bottomrule
\end{tabular}
\end{table*}

\subsection{What the Transformer attends to}
\begin{figure}[!htbp]
\centering
\includegraphics[width=\linewidth]{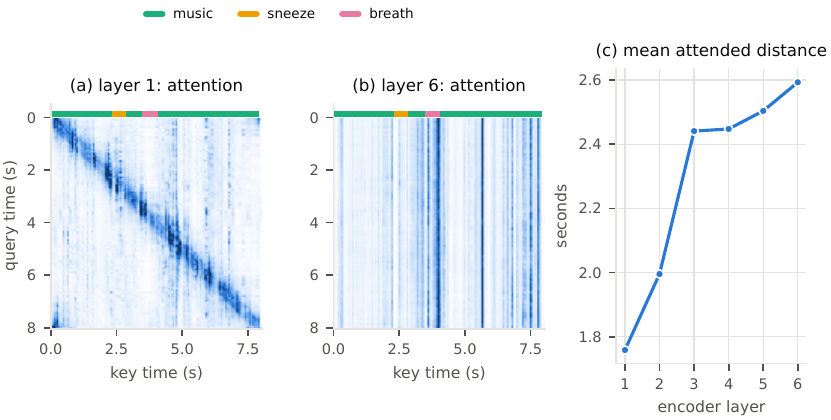}
\caption{Self-attention of encoder layers 1 (a) and 6 (b), averaged over the eight heads, for the clip of \cref{fig:example}; coloured bars on top mark the annotated events. (c) Mean attended distance per layer, $\sum_j a_{ij}\,|i-j|\cdot 40$\,ms averaged over queries $i$.}
\label{fig:attention}
\end{figure}

\Cref{fig:attention} shows the self-attention weights of the first and last encoder layers, averaged over heads, for the clip in \cref{fig:example} (one 8\,s window). The first layer attends mostly to nearby frames, which appears as a diagonal band: 54\,\% of its attention falls within $\pm1$\,s of the query. The last layer puts more weight on a few time positions that all queries share, visible as vertical stripes: its five most-attended positions receive 12.7\,\% of the attention, against 2.5\,\% for uniform attention. Over the six layers the mean attended distance grows from 1.76\,s to 2.59\,s (2.68\,s for uniform attention), so deeper layers combine information from a wider part of the window. The mean attention entropy per query is between 4.86 and 5.03\,nats, against $\ln 201 = 5.30$ for uniform attention. This is a single clip, and we do not claim that the attended positions correspond to particular events.

\subsection{Worked examples}
\label{sec:examples-eval}
\begin{figure*}[!htbp]
\centering
\includegraphics[width=0.8\textwidth]{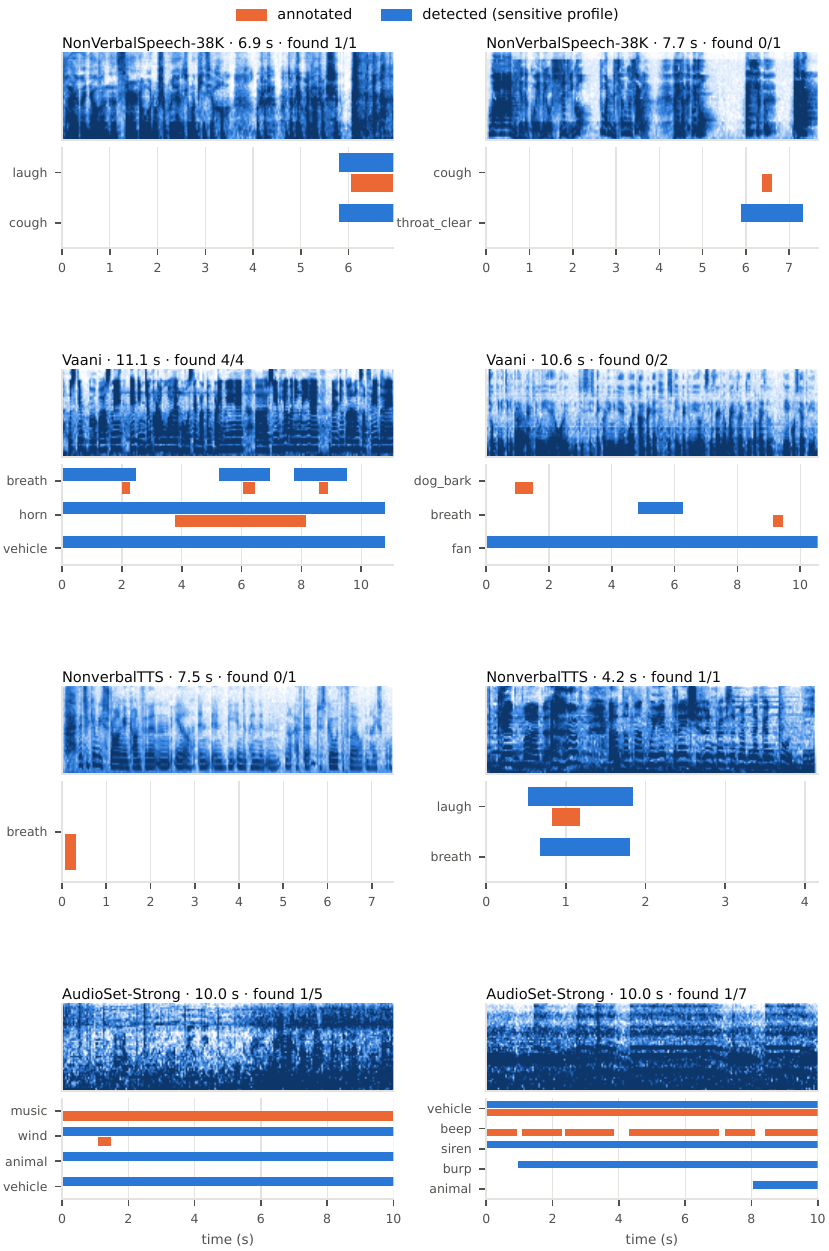}
\caption{Worked examples: eight held-out clips, two per source, chosen by a fixed rule (text). Each panel shows the log-mel spectrogram, the annotated events (orange, upper bar of each row) and the events detected with the sensitive profile (blue, lower bar). ``found'' counts annotated events overlapped by a same-class detection.}
\label{fig:examples}
\end{figure*}

To show typical behaviour beyond averages, \cref{fig:examples,tab:examples} give eight held-out clips, two per source. The selection is mechanical: for each source, the first two held-out clips in storage order that last 5--12\,s (NonverbalTTS: 4--12\,s; AudioSet-Strong: 9--11\,s, i.e.\ its 10\,s clips) and carry at least one annotated event (Vaani and AudioSet-Strong: at least two event classes), skipping the clip of \cref{fig:example}. Across the eight clips, the sensitive profile finds 8 of the 22 annotated events. It finds all four events of the first Vaani clip (a horn and three breaths) and both laughs, but misses a short cough, a dog bark and two breaths, does not detect the music in the first AudioSet clip (it finds only the short wind gust there), and finds none of the six short beeps in the second AudioSet clip, where its one hit is the annotated vehicle sound and it adds \evt{siren}, \evt{burp} and \evt{animal}. For the English NonverbalTTS clips, the merged Whisper transcripts are:
\begin{itemize}
\item \texttt{It's good, [laugh] but I [breath] have to go to Tray and go to Gantur.} (annotated: a laugh at 0.8--1.2\,s; the breath tag is not annotated, so under our protocol it counts as a false alarm, although it may be an unlabelled breath);
\item \texttt{That you can be strong and confident and you know and be independent and not have to just go with the flow in high school} (annotated: a breath at 0.1--0.3\,s, missed).
\end{itemize}
The transcripts are Whisper's own; the recognizer's word errors (e.g.\ place names) are not corrected.

\begin{table*}[!htbp]
\centering\scriptsize
\caption{Worked examples on held-out clips (selection rule in \cref{sec:examples-eval}). Times are in seconds. ``Found'' counts annotated events overlapped by a same-class detection, for the sensitive (S) and balanced (B) profiles. Detected classes that the clip's source does not annotate cannot be judged.}
\label{tab:examples}
\begin{tabularx}{\textwidth}{@{}l r >{\raggedright\arraybackslash}X >{\raggedright\arraybackslash}X cc@{}}
\toprule
Clip & Dur. & Annotated events & Detected (sensitive) & Found S & Found B \\
\midrule
nv-1 & 6.9 & \evt{laugh} 6.1--6.9 & \evt{cough} 5.8--7.0, \evt{laugh} 5.8--7.0 & 1/1 & 0/1 \\
nv-2 & 7.7 & \evt{cough} 6.4--6.6 & \evt{throat\_clear} 5.9--7.3 & 0/1 & 0/1 \\
vaani-1 & 11.1 & \evt{breath} 2.0--2.3, \evt{horn} 3.8--8.2, \evt{breath} 6.1--6.4, \evt{breath} 8.6--8.9 & \evt{breath} 0.0--2.5, \evt{horn} 0.0--10.8, \evt{vehicle} 0.0--10.8, \evt{breath} 5.2--7.0, \evt{breath} 7.8--9.5 & 4/4 & 4/4 \\
vaani-2 & 10.6 & \evt{dog\_bark} 0.9--1.5, \evt{breath} 9.1--9.4 & \evt{fan} 0.0--10.6, \evt{breath} 4.8--6.3 & 0/2 & 0/2 \\
nvtts-1 & 7.5 & \evt{breath} 0.1--0.3 & -- & 0/1 & 0/1 \\
nvtts-2 & 4.2 & \evt{laugh} 0.8--1.2 & \evt{laugh} 0.5--1.8, \evt{breath} 0.7--1.8 & 1/1 & 0/1 \\
audioset-1 & 10.0 & \evt{music} 0.0--10.0, \evt{wind} 1.1--1.5, \evt{music} 3.0--4.5, \evt{music} 4.7--6.3, \evt{music} 8.9--9.6 & \evt{animal} 0.0--10.0, \evt{vehicle} 0.0--10.0, \evt{wind} 0.0--10.0 & 1/5 & 0/5 \\
audioset-2 & 10.0 & \evt{vehicle} 0.0--10.0, \evt{beep} 0.0--0.9, \evt{beep} 1.1--2.3, \evt{beep} 2.4--3.9, \evt{beep} 4.3--7.0, \evt{beep} 7.2--8.1, \evt{beep} 8.4--10.0 & \evt{siren} 0.0--10.0, \evt{vehicle} 0.0--10.0, \evt{burp} 1.0--10.0, \evt{animal} 8.0--10.0 & 1/7 & 0/7 \\
\bottomrule
\end{tabularx}
\end{table*}

\subsection{Qualitative example}
\begin{figure}[!htbp]
\centering
\includegraphics[width=\linewidth]{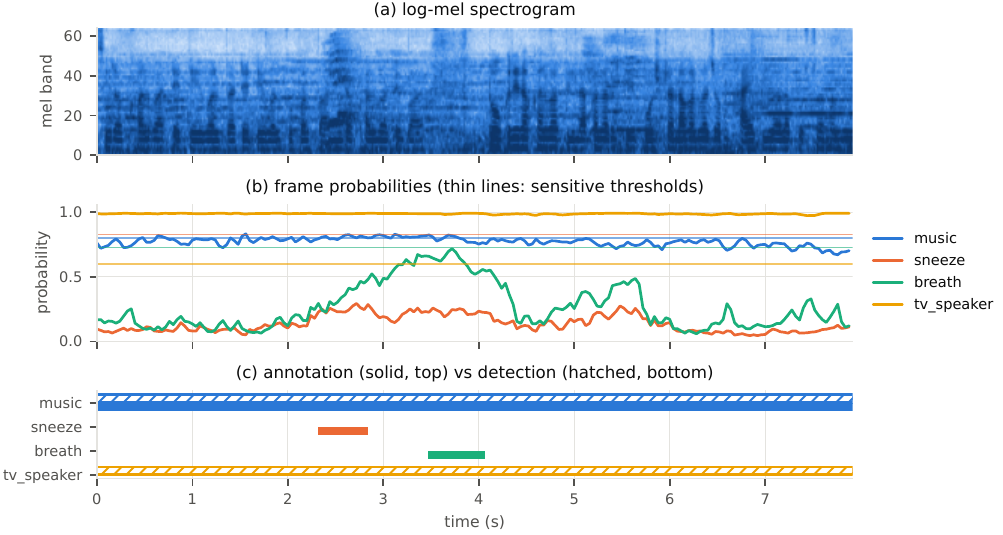}
\caption{A held-out Vaani clip (Deoria district; selection rule in the text): (a) log-mel spectrogram; (b) frame probabilities of the annotated and detected classes, with their sensitive thresholds as thin lines; (c) annotated events (solid) and detected events (hatched). Music is detected; the sneeze and the breath are missed; \evt{tv\_speaker} is a false alarm.}
\label{fig:example}
\end{figure}

\Cref{fig:example} shows a held-out Vaani clip. To avoid hand-picking, the clip is the first held-out Vaani clip in storage order that contains at least three event classes and lasts 6--12\,s. It is annotated with music throughout, a sneeze at 2.3\,s and a breath at 3.5\,s. The model detects the music, but misses the sneeze and the breath, whose probabilities stay below their thresholds (the breath peaks at about 0.7), and adds a false \evt{tv\_speaker}, a weak class, throughout. A plausible reason is that music under speech recorded on a phone resembles the TV and radio recordings that dominate \evt{tv\_speaker}'s training data (\cref{fig:crops}), but a single clip cannot establish this. The example shows two kinds of error that also occur in \cref{fig:examples}: confusions among stationary background classes, and short vocal events just below their thresholds.

In an informal check on a 6.8\,s English recording made by the first author, with a sigh at the beginning and a cough at about 4\,s, the model (sensitive profile) tagged the sigh (0.00--1.04\,s) and the cough (3.64--4.88\,s), together with a \evt{throat\_clear} overlapping the cough and a spurious \evt{yawn}. This single example is anecdotal and is not part of the evaluation.

\subsection{Speed}
\label{sec:speed}
Detection of 178\,s of audio, excluding audio decoding and model loading, took 4.6\,s on eight CPU threads (39$\times$ real time) and 0.47\,s on a T4 GPU (380$\times$ real time), measured with the released code (\texttt{Detector.detect}). Transcription with Whisper, when used, adds its own run time, which we did not measure.

\section{Limitations, licensing and ethics}
\label{sec:ethics}

\paragraph{Evaluation.} The caveats of \cref{sec:protocol} limit what our numbers show: checkpoint selection and thresholds used validation windows from the same held-out clips as the evaluation, there is a single run, no standard benchmark was used, and our event criterion ignores boundary accuracy. The NonverbalTTS timestamps (\cref{sec:tts}) and the unverified Vaani timestamps used for training are not manually checked. The most informative next step is a small, carefully labelled test set from the target domain.

\paragraph{Domain coverage.} Most real training audio is Mandarin radio drama and comedy (NVS-38K, by its own source statistics~\citep{nvs38k}), Indic speech recorded on phones (Vaani), English interviews and acted speech (NonverbalTTS) and YouTube audio (AudioSet-Strong). Accuracy on other domains, such as call-centre, meeting or far-field audio, is unknown. Eleven classes are weak (\cref{tab:perclass}), and steady background noise can trigger \evt{fan} (as in the second Vaani clip of \cref{fig:examples}).

\paragraph{Design choices not explored.} We did not try pretrained audio or speech encoders such as BEATs~\citep{chen2023beats} or WavLM~\citep{chen2022wavlm}, other window lengths, or other loss weightings, and we did not ablate the individual data sources, augmentations or the bed cleaning. The contribution of each component is therefore not isolated.

\paragraph{Licensing.} Three training sources restrict use: NVS-38K is licensed CC BY-NC 4.0 and limited to non-commercial research; the NonverbalTTS annotations are CC BY-NC-SA 4.0, with the audio under the terms of VoxCeleb and Expresso; OpenSLR\,99 is CC BY-NC-ND 4.0. The AudioSet-Strong mirror states no license, and AudioSet audio originates from YouTube. Whether trained weights are derivative works of their training data is not settled. We therefore release the weights for non-commercial use only (CC BY-NC-SA 4.0); a version for other uses would have to be retrained without the restricted sources. OpenSLR\,99 contributed only 591 of about 138{,}000 event crops.

\paragraph{Intended use and risks.} The model is intended for rich transcription, corpus analysis, data filtering and annotation assistance with human review. It must not be used to infer health conditions (for example from coughs), for surveillance, or for decisions about individuals. The training data contains recordings of real people; we use it as published by its creators and redistribute no audio.

\section{Conclusion}
\label{sec:conclusion}
We built a compact detector for 54 non-verbal and background sound classes with 40\,ms resolution by unifying seven public datasets with incompatible label formats, recovering missing timestamps by forced alignment, training on a mix of real and synthetic windows with real-noise augmentation, and calibrating per-class thresholds into two profiles on real held-out audio. On held-out real recordings, the sensitive profile finds two thirds of the annotated events of the 48 classes with at least ten held-out events (with thresholds tuned on overlapping held-out data), and the model runs far faster than real time on a CPU. The main open problems are an evaluation on a properly separated, target-domain test set, the weak classes, and the licensing constraints inherited from the training data.

\clearpage
\bibliographystyle{plainnat}
\bibliography{references}

\appendix
\section{Per-class results}
\label{app:perclass}
\begingroup\footnotesize\setlength{\tabcolsep}{3.2pt}
\begin{longtable}{@{}llcrrrrrrrrrrr@{}}
\caption{Per-class results. Type: S = speaker sound, B = background. Eval: validation sources that label the class (N = NVS-38K, V = Vaani, T = NonverbalTTS, A = AudioSet-Strong). Pos.\ = positive validation frames (40\,ms). AP = frame-level average precision. $\theta$ = threshold; CV = 2-fold cross-validated frame F1 (balanced) or F2 (sensitive). Event level: held-out events, and event recall (R) and precision (P) with the sensitive (S) and balanced (B) profiles. $^\dagger$ weak class.}
\label{tab:perclass}\\
\toprule
 & & & & & \multicolumn{2}{c}{balanced} & \multicolumn{2}{c}{sensitive} & \multicolumn{5}{c}{event level} \\
\cmidrule(lr){6-7}\cmidrule(lr){8-9}\cmidrule(l){10-14}
Class & Type & Eval & Pos. & AP & $\theta$ & CV F1 & $\theta$ & CV F2 & Events & R-S & P-S & R-B & P-B \\
\midrule
\endfirsthead
\toprule
Class & Type & Eval & Pos. & AP & $\theta$ & CV F1 & $\theta$ & CV F2 & Events & R-S & P-S & R-B & P-B \\
\midrule
\endhead
\bottomrule
\endfoot
\texttt{breath} & S & NVTA & 16{,}762 & 0.499 & 0.850 & 0.499 & 0.725 & 0.616 & 1{,}116 & 0.80 & 0.49 & 0.67 & 0.64 \\
\texttt{burp} & S & A & 136 & 0.218 & 0.650 & 0.285 & 0.375 & 0.361 & 7 & 0.57 & 0.16 & 0.29 & 0.10 \\
\texttt{chew}$^\dagger$ & S & A & 242 & 0.039 & 0.900 & 0.056 & 0.600 & 0.108 & 29 & 0.10 & 0.08 & 0.00 & 0.00 \\
\texttt{cough} & S & NVTA & 2{,}033 & 0.313 & 0.875 & 0.400 & 0.750 & 0.493 & 329 & 0.56 & 0.44 & 0.39 & 0.59 \\
\texttt{cry}$^\dagger$ & S & NA & 963 & 0.075 & 0.900 & 0.048 & 0.600 & 0.110 & 61 & 0.43 & 0.11 & 0.08 & 0.09 \\
\texttt{gasp} & S & NVA & 268 & 0.065 & 0.775 & 0.126 & 0.750 & 0.160 & 41 & 0.37 & 0.12 & 0.34 & 0.13 \\
\texttt{groan}$^\dagger$ & S & TA & 293 & 0.015 & 0.900 & 0.028 & 0.600 & 0.050 & 36 & 0.19 & 0.09 & 0.11 & 0.14 \\
\texttt{hiccup}$^\dagger$ & S & A & 73 & 0.004 & 0.900 & 0.015 & 0.600 & 0.034 & 14 & 0.21 & 0.60 & 0.00 & 0.00 \\
\texttt{laugh} & S & NVTA & 5{,}947 & 0.516 & 0.950 & 0.467 & 0.725 & 0.561 & 587 & 0.79 & 0.52 & 0.57 & 0.77 \\
\texttt{lip\_smack} & S & V & 168 & 0.052 & 0.925 & 0.126 & 0.850 & 0.135 & 9 & 0.22 & 0.15 & 0.22 & 0.25 \\
\texttt{nose\_blow}$^\dagger$ & S & V & 36 & 0.000 & 0.900 & 0.000 & 0.600 & 0.000 & 1 & 0.00 & 0.00 & 0.00 & 0.00 \\
\texttt{scream} & S & A & 503 & 0.423 & 0.950 & 0.133 & 0.950 & 0.235 & 32 & 0.16 & 0.50 & 0.16 & 0.50 \\
\texttt{shout} & S & A & 1{,}783 & 0.147 & 0.800 & 0.258 & 0.725 & 0.371 & 97 & 0.45 & 0.19 & 0.35 & 0.19 \\
\texttt{sigh} & S & NTA & 2{,}909 & 0.725 & 0.975 & 0.652 & 0.900 & 0.718 & 389 & 0.84 & 0.72 & 0.60 & 0.86 \\
\texttt{sneeze} & S & VTA & 144 & 0.429 & 0.975 & 0.212 & 0.825 & 0.406 & 8 & 0.00 & 0.00 & 0.00 & 0.00 \\
\texttt{sniff} & S & NVTA & 1{,}368 & 0.438 & 0.925 & 0.421 & 0.875 & 0.526 & 272 & 0.65 & 0.60 & 0.56 & 0.71 \\
\texttt{snore} & S & NVTA & 272 & 0.551 & 0.975 & 0.610 & 0.950 & 0.567 & 28 & 0.68 & 0.43 & 0.68 & 0.46 \\
\texttt{throat\_clear} & S & NVTA & 1{,}024 & 0.335 & 0.900 & 0.383 & 0.800 & 0.471 & 190 & 0.62 & 0.38 & 0.45 & 0.49 \\
\texttt{whisper} & S & A & 215 & 0.297 & 0.975 & 0.374 & 0.950 & 0.311 & 36 & 0.36 & 0.83 & 0.28 & 0.75 \\
\texttt{yawn}$^\dagger$ & S & NVA & 122 & 0.007 & 0.900 & 0.014 & 0.600 & 0.033 & 14 & 0.07 & 0.01 & 0.00 & 0.00 \\
\texttt{animal} & B & VA & 6{,}034 & 0.151 & 0.800 & 0.139 & 0.375 & 0.178 & 538 & 0.52 & 0.12 & 0.25 & 0.25 \\
\texttt{applause} & B & A & 1{,}785 & 0.582 & 0.925 & 0.480 & 0.900 & 0.499 & 199 & 0.34 & 0.47 & 0.27 & 0.47 \\
\texttt{baby\_cry} & B & VA & 779 & 0.218 & 0.975 & 0.402 & 0.950 & 0.436 & 31 & 0.39 & 0.23 & 0.29 & 0.21 \\
\texttt{beep} & B & VA & 2{,}025 & 0.086 & 0.725 & 0.184 & 0.700 & 0.255 & 165 & 0.56 & 0.26 & 0.52 & 0.27 \\
\texttt{bell} & B & VA & 3{,}657 & 0.261 & 0.975 & 0.347 & 0.775 & 0.402 & 87 & 0.81 & 0.19 & 0.41 & 0.36 \\
\texttt{bird} & B & VA & 22{,}291 & 0.399 & 0.925 & 0.426 & 0.800 & 0.535 & 791 & 0.64 & 0.60 & 0.56 & 0.80 \\
\texttt{buzzer} & B & VA & 845 & 0.089 & 0.775 & 0.118 & 0.450 & 0.215 & 25 & 0.20 & 0.06 & 0.16 & 0.17 \\
\texttt{cheer} & B & A & 1{,}997 & 0.321 & 0.900 & 0.331 & 0.875 & 0.408 & 73 & 0.26 & 0.39 & 0.23 & 0.43 \\
\texttt{child\_voice} & B & VA & 4{,}804 & 0.149 & 0.950 & 0.188 & 0.775 & 0.335 & 179 & 0.48 & 0.22 & 0.25 & 0.39 \\
\texttt{construction} & B & A & 3{,}147 & 0.213 & 0.800 & 0.329 & 0.525 & 0.272 & 220 & 0.67 & 0.14 & 0.42 & 0.22 \\
\texttt{crowd} & B & A & 2{,}484 & 0.320 & 0.975 & 0.238 & 0.850 & 0.247 & 41 & 0.61 & 0.17 & 0.41 & 0.34 \\
\texttt{dishes} & B & A & 324 & 0.196 & 0.650 & 0.202 & 0.625 & 0.242 & 50 & 0.42 & 0.11 & 0.38 & 0.09 \\
\texttt{dog\_bark} & B & VA & 3{,}209 & 0.507 & 0.975 & 0.475 & 0.650 & 0.514 & 128 & 0.66 & 0.17 & 0.45 & 0.77 \\
\texttt{door}$^\dagger$ & B & A & 162 & 0.076 & 0.900 & 0.031 & 0.600 & 0.067 & 21 & 0.05 & 0.08 & 0.00 & 0.00 \\
\texttt{explosion} & B & A & 911 & 0.224 & 0.975 & 0.460 & 0.975 & 0.379 & 54 & 0.63 & 0.16 & 0.63 & 0.16 \\
\texttt{fan}$^\dagger$ & B & VA & 1{,}725 & 0.035 & 0.900 & 0.040 & 0.800 & 0.087 & 13 & 0.15 & 0.03 & 0.08 & 0.03 \\
\texttt{footsteps}$^\dagger$ & B & A & 1{,}185 & 0.041 & 0.900 & 0.077 & 0.650 & 0.137 & 218 & 0.50 & 0.10 & 0.30 & 0.29 \\
\texttt{glass\_break} & B & A & 450 & 0.200 & 0.975 & 0.210 & 0.975 & 0.193 & 58 & 0.12 & 1.00 & 0.12 & 1.00 \\
\texttt{gunshot} & B & A & 560 & 0.308 & 0.975 & 0.323 & 0.950 & 0.311 & 75 & 0.81 & 0.22 & 0.79 & 0.37 \\
\texttt{horn} & B & VA & 10{,}844 & 0.590 & 0.975 & 0.557 & 0.900 & 0.604 & 258 & 0.76 & 0.71 & 0.64 & 0.83 \\
\texttt{insect} & B & VA & 5{,}945 & 0.380 & 0.975 & 0.430 & 0.925 & 0.488 & 87 & 0.42 & 0.23 & 0.39 & 0.35 \\
\texttt{knock}$^\dagger$ & B & A & 0 & 0.000 & 0.900 & 0.000 & 0.600 & 0.000 & 5 & 0.00 & 0.00 & 0.00 & 0.00 \\
\texttt{machine} & B & VA & 3{,}273 & 0.138 & 0.800 & 0.276 & 0.750 & 0.296 & 137 & 0.51 & 0.12 & 0.45 & 0.12 \\
\texttt{music} & B & VA & 60{,}578 & 0.884 & 0.925 & 0.809 & 0.800 & 0.831 & 1{,}361 & 0.84 & 0.76 & 0.75 & 0.85 \\
\texttt{phone\_ring} & B & VA & 1{,}263 & 0.086 & 0.875 & 0.196 & 0.850 & 0.245 & 64 & 0.70 & 0.24 & 0.67 & 0.23 \\
\texttt{phone\_vibrate} & B & VA & 139 & 0.165 & 0.800 & 0.401 & 0.800 & 0.386 & 4 & 0.25 & 0.09 & 0.25 & 0.09 \\
\texttt{rain} & B & A & 1{,}553 & 0.160 & 0.950 & 0.282 & 0.950 & 0.350 & 43 & 0.12 & 0.07 & 0.12 & 0.07 \\
\texttt{siren} & B & VA & 2{,}575 & 0.490 & 0.975 & 0.475 & 0.950 & 0.505 & 114 & 0.58 & 0.24 & 0.50 & 0.29 \\
\texttt{thunder} & B & A & 345 & 0.104 & 0.975 & 0.187 & 0.925 & 0.228 & 13 & 0.54 & 0.17 & 0.31 & 0.18 \\
\texttt{tv\_speaker}$^\dagger$ & B & VA & 2{,}284 & 0.044 & 0.925 & 0.069 & 0.600 & 0.169 & 23 & 0.43 & 0.06 & 0.22 & 0.09 \\
\texttt{vehicle} & B & VA & 23{,}774 & 0.497 & 0.850 & 0.494 & 0.725 & 0.563 & 549 & 0.87 & 0.32 & 0.80 & 0.44 \\
\texttt{water} & B & A & 7{,}584 & 0.385 & 0.900 & 0.380 & 0.625 & 0.470 & 192 & 0.75 & 0.19 & 0.49 & 0.27 \\
\texttt{whistle} & B & VA & 1{,}593 & 0.521 & 0.950 & 0.492 & 0.850 & 0.536 & 83 & 0.28 & 0.38 & 0.17 & 0.44 \\
\texttt{wind} & B & A & 7{,}780 & 0.316 & 0.950 & 0.297 & 0.850 & 0.357 & 206 & 0.68 & 0.17 & 0.45 & 0.26 \\
\end{longtable}
\endgroup

\section{Annotated events per class and source}
\label{app:counts}
\begingroup\footnotesize\setlength{\tabcolsep}{4pt}
\begin{longtable}{@{}llrrrrrrrr@{}}
\caption{Annotated events per class in each source's full index, before any sampling caps. DEMAND counts environments.}
\label{tab:counts}\\
\toprule
Class & Type & NVS-38K & Vaani & NV-TTS & AudioSet & SLR99 & DNC & DEMAND & Total \\
\midrule
\endfirsthead
\toprule
Class & Type & NVS-38K & Vaani & NV-TTS & AudioSet & SLR99 & DNC & DEMAND & Total \\
\midrule
\endhead
\bottomrule
\endfoot
\texttt{breath} & S & 2{,}162 & 35{,}663 & 6{,}121 & 27{,}232 & 44 & 0 & 0 & 71{,}222 \\
\texttt{music} & B & 0 & 6{,}872 & 0 & 51{,}454 & 0 & 229 & 0 & 58{,}555 \\
\texttt{bird} & B & 0 & 17{,}694 & 0 & 37{,}443 & 0 & 0 & 0 & 55{,}137 \\
\texttt{vehicle} & B & 0 & 4{,}854 & 0 & 25{,}450 & 0 & 692 & 4 & 31{,}000 \\
\texttt{wind} & B & 0 & 0 & 0 & 25{,}079 & 0 & 0 & 0 & 25{,}079 \\
\texttt{laugh} & S & 6{,}912 & 99 & 1{,}278 & 14{,}489 & 25 & 0 & 0 & 22{,}803 \\
\texttt{horn} & B & 0 & 15{,}739 & 0 & 3{,}794 & 0 & 0 & 0 & 19{,}533 \\
\texttt{animal} & B & 0 & 1{,}577 & 0 & 15{,}820 & 0 & 0 & 0 & 17{,}397 \\
\texttt{child\_voice} & B & 0 & 3{,}165 & 0 & 11{,}447 & 0 & 0 & 0 & 14{,}612 \\
\texttt{footsteps} & B & 0 & 0 & 0 & 12{,}490 & 0 & 0 & 0 & 12{,}490 \\
\texttt{dog\_bark} & B & 0 & 2{,}005 & 0 & 10{,}429 & 0 & 0 & 0 & 12{,}434 \\
\texttt{sigh} & S & 9{,}783 & 0 & 147 & 206 & 51 & 0 & 0 & 10{,}187 \\
\texttt{applause} & B & 0 & 0 & 0 & 10{,}157 & 0 & 0 & 0 & 10{,}157 \\
\texttt{water} & B & 0 & 0 & 0 & 10{,}053 & 0 & 0 & 0 & 10{,}053 \\
\texttt{beep} & B & 0 & 1{,}039 & 0 & 8{,}435 & 0 & 0 & 0 & 9{,}474 \\
\texttt{machine} & B & 0 & 156 & 0 & 7{,}379 & 0 & 853 & 1 & 8{,}389 \\
\texttt{cough} & S & 5{,}940 & 202 & 272 & 1{,}818 & 50 & 0 & 0 & 8{,}282 \\
\texttt{sniff} & S & 7{,}321 & 32 & 407 & 473 & 0 & 0 & 0 & 8{,}233 \\
\texttt{insect} & B & 0 & 3{,}324 & 0 & 4{,}561 & 0 & 0 & 0 & 7{,}885 \\
\texttt{shout} & S & 0 & 0 & 0 & 6{,}244 & 0 & 0 & 0 & 6{,}244 \\
\texttt{siren} & B & 0 & 129 & 0 & 6{,}023 & 0 & 0 & 0 & 6{,}152 \\
\texttt{bell} & B & 0 & 668 & 0 & 4{,}934 & 0 & 0 & 0 & 5{,}602 \\
\texttt{throat\_clear} & S & 4{,}909 & 96 & 214 & 278 & 54 & 0 & 0 & 5{,}551 \\
\texttt{construction} & B & 0 & 0 & 0 & 5{,}400 & 0 & 0 & 0 & 5{,}400 \\
\texttt{gunshot} & B & 0 & 0 & 0 & 5{,}074 & 0 & 0 & 0 & 5{,}074 \\
\texttt{whistle} & B & 0 & 106 & 0 & 4{,}061 & 0 & 0 & 0 & 4{,}167 \\
\texttt{dishes} & B & 0 & 0 & 0 & 3{,}654 & 0 & 0 & 0 & 3{,}654 \\
\texttt{baby\_cry} & B & 0 & 1{,}620 & 0 & 1{,}576 & 0 & 0 & 0 & 3{,}196 \\
\texttt{crowd} & B & 0 & 0 & 0 & 3{,}130 & 0 & 0 & 0 & 3{,}130 \\
\texttt{cheer} & B & 0 & 0 & 0 & 3{,}118 & 0 & 0 & 0 & 3{,}118 \\
\texttt{tv\_speaker} & B & 0 & 156 & 0 & 1{,}804 & 0 & 1{,}100 & 0 & 3{,}060 \\
\texttt{phone\_ring} & B & 0 & 1{,}548 & 0 & 1{,}409 & 0 & 0 & 0 & 2{,}957 \\
\texttt{explosion} & B & 0 & 0 & 0 & 2{,}505 & 0 & 0 & 0 & 2{,}505 \\
\texttt{cry} & S & 753 & 0 & 0 & 1{,}234 & 24 & 0 & 0 & 2{,}011 \\
\texttt{gasp} & S & 578 & 1 & 0 & 1{,}166 & 0 & 0 & 0 & 1{,}745 \\
\texttt{chew} & S & 0 & 0 & 0 & 1{,}730 & 0 & 0 & 0 & 1{,}730 \\
\texttt{rain} & B & 0 & 0 & 0 & 1{,}711 & 0 & 0 & 0 & 1{,}711 \\
\texttt{lip\_smack} & S & 0 & 1{,}437 & 0 & 0 & 99 & 0 & 0 & 1{,}536 \\
\texttt{whisper} & S & 0 & 0 & 0 & 1{,}470 & 0 & 0 & 0 & 1{,}470 \\
\texttt{scream} & S & 0 & 0 & 0 & 1{,}330 & 51 & 0 & 0 & 1{,}381 \\
\texttt{snore} & S & 91 & 12 & 15 & 1{,}101 & 0 & 0 & 0 & 1{,}219 \\
\texttt{glass\_break} & B & 0 & 0 & 0 & 1{,}159 & 0 & 0 & 0 & 1{,}159 \\
\texttt{groan} & S & 0 & 0 & 148 & 901 & 45 & 0 & 0 & 1{,}094 \\
\texttt{fan} & B & 0 & 599 & 0 & 452 & 0 & 0 & 0 & 1{,}051 \\
\texttt{buzzer} & B & 0 & 76 & 0 & 839 & 0 & 0 & 0 & 915 \\
\texttt{door} & B & 0 & 0 & 0 & 895 & 0 & 0 & 0 & 895 \\
\texttt{burp} & S & 0 & 0 & 0 & 765 & 0 & 0 & 0 & 765 \\
\texttt{thunder} & B & 0 & 0 & 0 & 728 & 0 & 0 & 0 & 728 \\
\texttt{hiccup} & S & 0 & 0 & 0 & 608 & 0 & 0 & 0 & 608 \\
\texttt{sneeze} & S & 0 & 87 & 18 & 440 & 52 & 0 & 0 & 597 \\
\texttt{knock} & B & 0 & 0 & 0 & 431 & 0 & 0 & 0 & 431 \\
\texttt{phone\_vibrate} & B & 0 & 231 & 0 & 176 & 0 & 0 & 0 & 407 \\
\texttt{yawn} & S & 269 & 36 & 0 & 19 & 48 & 0 & 0 & 372 \\
\texttt{nose\_blow} & S & 0 & 97 & 0 & 0 & 48 & 0 & 0 & 145 \\
\end{longtable}
\endgroup

\section{Label-mapping rules}
\label{app:rules}
The rules below are reproduced verbatim from the training pipeline (\texttt{training/build\_notebooks.py}), except that the one-to-one NVS-38K map (\cref{sec:labels}) and the NonverbalTTS emoji map are omitted; for each rule list, the first matching pattern wins. \texttt{None} means ignored, and \texttt{DROP} (in \texttt{canon\_vaani}) means a tag that makes the clip unusable as a real window. The NonverbalTTS emoji map is omitted here because it consists of pictographs; it maps breath, laughing, sniffing, coughing, throat clearing, sighing, groaning, sneezing, snoring and grunting to the classes of \cref{sec:labels}, with grunting mapped to \evt{groan}.
\begin{lstlisting}[language=Python]
# Label-mapping rules used to train the released model (verbatim from training/build_notebooks.py).
VAANI_RULES = [   # first match wins, so order matters
    (r"\b(child|children|kids?|baby|human|people)\b.*\b(talking|speaking|voices?)\b|human (noise|sound)", None),
    (r"\b(child|children|kids?|baby)\b.*\bcry",                        "baby_cry"),
    (r"laugh",                                                         "laugh"),
    (r"cough",                                                         "cough"),
    (r"throat",                                                        "throat_clear"),
    (r"sneez",                                                         "sneeze"),
    (r"nose (blow|clear)",                                             "nose_blow"),
    (r"lip.?smack",                                                    "lip_smack"),
    (r"breath|inhal|exhal",                                            "breath"),
    (r"sniff|snort",                                                   "sniff"),
    (r"yawn|yawing",                                                   "yawn"),
    (r"snor",                                                          "snore"),
    (r"gasp",                                                          "gasp"),
    (r"\b(child|children|kids?|baby)\b",                               "child_voice"),   # yelling / playing / noise
    (r"siren|ambulance|alarm",                                         "siren"),
    (r"horn|honk",                                                     "horn"),
    (r"bark|\bdog\b",                                                  "dog_bark"),
    (r"insect|cricket|\bbugs?\b|beetle",                               "insect"),
    (r"bird|crow|caw|rooster|\bcock\b|\bhen\b|cluck|duck|cuckoo|sparrow|squawk|twitter|chirp", "bird"),
    (r"\bcow\b|goat|\bcat\b|meow|buffalo|frog|squirrel|animal",        "animal"),
    (r"whistl",                                                        "whistle"),
    (r"music|song|sing|flute|drum|band sound|prayer|mantra|chant|record playing", "music"),
    (r"\bfan\b",                                                       "fan"),           # before "ring": fan whiRRING
    (r"vibrat",                                                        "phone_vibrate"),
    (r"bell|chim",                                                     "bell"),          # before "ring": bell ringing
    (r"\bring(ing|tone|s)?\b|ringtone",                                "phone_ring"),
    (r"beep|notification|message tone|phone sound|mobile sound|phone click", "beep"),
    (r"buzz",                                                          "buzzer"),
    (r"\btv\b|television|speaker|\bmic\b|\bmike\b",                    "tv_speaker"),
    (r"vehicle|traffic|engine|bike|motor|\bcar\b|train|tractor|aeroplane|rickshaw", "vehicle"),
    (r"machine|generator|factory|typing|tick|clock|pump",              "machine"),
]
_RULES = [(re.compile(p), c) for p, c in VAANI_RULES]
DROP = "__drop__"
MIN_EVENTS = 80

def canon_vaani(tag):
    t = re.sub(r"\s+", " ", re.sub(r"[<>\[\]_]", " ", str(tag).lower())).strip()
    for rx, c in _RULES:
        if rx.search(t):
            return c
    return DROP

# (NonverbalTTS emoji map omitted here; see Section 4.1 of the paper and sed/taxonomy.py)

SLR_MAP = {"coughing": "cough", "yawning": "yawn", "throat-clearing": "throat_clear", "sighing": "sigh",
           "lip-smacking": "lip_smack", "lip-popping": "lip_smack", "panting": "breath", "crying": "cry",
           "laughing": "laugh", "sneezing": "sneeze", "nose-blowing": "nose_blow", "moaning": "groan",
           "screaming": "scream", "tongue-clicking": "click", "teeth-chattering": None, "teeth-grinding": None}

# AudioSet ontology names -> canonical tags (first match wins). Unmatched events (speech, impacts, ...) are ignored.
AS_RULES = [
    (r"speech synthesizer|howl \(wind\)|sliding door|sanding|bellow|crowd|growling", None),   # false friends
    (r"^(baby cry|infant cry)",                                         "baby_cry"),
    (r"baby laughter|^laughter|giggle|snicker|belly laugh|chuckle",     "laugh"),
    (r"^crying|sobbing|^whimper$",                                      "cry"),
    (r"^cough",                                                         "cough"),
    (r"^sneeze",                                                        "sneeze"),
    (r"^sniff",                                                         "sniff"),
    (r"^snoring",                                                       "snore"),
    (r"^sigh",                                                          "sigh"),
    (r"^gasp",                                                          "gasp"),
    (r"throat clearing",                                                "throat_clear"),
    (r"^yawn",                                                          "yawn"),
    (r"^(groan|grunt)",                                                 "groan"),
    (r"^(screaming|shriek|yell)",                                       "scream"),
    (r"^(breathing|pant|wheeze)",                                       "breath"),
    (r"^whispering",                                                    "whisper"),
    (r"^shout|battle cry",                                              "shout"),
    (r"^chewing",                                                       "chew"),
    (r"^burping",                                                       "burp"),
    (r"^hiccup",                                                        "hiccup"),
    (r"nose blowing",                                                   "nose_blow"),
    (r"lip smack",                                                      "lip_smack"),
    (r"^child|^children",                                               "child_voice"),
    (r"siren|emergency vehicle|alarm",                                  "siren"),
    (r"horn|honking|^toot$",                                             "horn"),
    (r"^(bark|dog|yip|bow-wow|howl|bay)",                               "dog_bark"),
    (r"insect|cricket|mosquito|^fly|bee, wasp|buzzing insect",          "insect"),
    (r"bird|chirp|tweet|^crow($|ing)|^caw|squawk|pigeon|^coo$|rooster|chicken|duck|quack|^owl", "bird"),
    (r"cattle|moo|goat|bleat|^cat$|meow|purr|frog|croak|horse|neigh|^pig|oink|sheep|animal", "animal"),
    (r"whistl",                                                         "whistle"),
    (r"television|^radio",                                              "tv_speaker"),
    (r"mechanical fan|air conditioning",                                "fan"),
    (r"vibrating alert",                                                "phone_vibrate"),
    (r"telephone bell|ringtone|^telephone$",                            "phone_ring"),
    (r"bell\b|^ding|wind chime",                                                 "bell"),
    (r"beep|bleep|dtmf",                                                "beep"),
    (r"^buzz",                                                          "buzzer"),
    (r"music|singing|song|choir|orchestra|guitar|piano|drum|flute|violin|synthesizer|keyboard \(musical\)|instrument", "music"),
    (r"vacuum cleaner|washing machine|dishwasher|blender|hair dryer|electric shaver|sewing machine|microwave|printer|typing|computer keyboard|clock|tick-tock", "machine"),
    (r"^(clapping|applause)",                                           "applause"),
    (r"^(cheering|whoop)",                                              "cheer"),
    (r"^crowd$|hubbub|speech babble",                                   "crowd"),
    (r"footsteps|^run$",                                                "footsteps"),
    (r"^wind$|^wind noise",                                                 "wind"),
    (r"thunder",                                                        "thunder"),
    (r"^rain|^patter",                                                  "rain"),
    (r"^water|splash|^pour|^drip|trickle|^stream$|^liquid|gurgling|toilet flush|waves|^ocean", "water"),
    (r"^knock",                                                         "knock"),
    (r"^door$|^slam",                                                   "door"),
    (r"^dishes|cutlery|chink, clink",                                   "dishes"),
    (r"^shatter|^breaking",                                     "glass_break"),
    (r"gunshot|machine gun|cap gun|artillery",                          "gunshot"),
    (r"explosion|firecracker|fireworks",                                "explosion"),
    (r"^hammer|power tool|^drill|chainsaw|^sawing|^tools$",             "construction"),
    (r"vehicle|^car|truck|^bus$|helicopter|motorboat|subway|air brake|motorcycle|traffic|engine|idling|revving|race car|train|railroad|rail transport|aircraft|tractor", "vehicle"),
]
_AS_RULES = [(re.compile(p), c) for p, c in AS_RULES]
def canon_as(name):
    t = str(name).lower()
    for rx, c in _AS_RULES:
        if rx.search(t): return c
    return None

# Noise datasets: labelled ambient sounds, or NOISE = unlabelled background for the augmentation pool.
NOISE = "__noise__"
DNC_MAP = {"TV": "tv_speaker", "Radio": "tv_speaker", "Music": "music", "autos": "vehicle",
           "WashingMachine": "machine", "WaterHeater": "machine", "Dishwasher": "machine", "CoffeeMachine": "machine",
           "VacuumCleaner": "machine", "Microwave": "machine", "quiet": NOISE, "people": NOISE}
DEMAND_MAP = {"STRAFFIC": "vehicle", "TBUS": "vehicle", "TCAR": "vehicle", "TMETRO": "vehicle", "DWASHING": "machine"}
                                                   # every other DEMAND environment -> NOISE

\end{lstlisting}

\section{Reproducibility}
\label{app:repro}
\paragraph{Released material.} The code repository contains the inference library (\texttt{sed/}), the command-line tool, the two training notebooks and their generator, the trained weights (\texttt{best.pt} and \texttt{model.safetensors} with \texttt{config.json}), both threshold files, the post-processing configuration, the per-class threshold report, the event-level results and the per-epoch training log. This report's folder contains every number used here (\texttt{data/}), the scripts that turn them into the figures and tables (\texttt{scripts/make\_figures.py}, \texttt{scripts/make\_tables.py}), the scripts that recompute the dataset statistics and summaries (\texttt{scripts/compute\_stats.py}, \texttt{scripts/compute\_summary.py}), and the scripts that extracted the example clips, layer table, activations and attention maps from the model (\texttt{scripts/prepare\_example.py}, \texttt{scripts/prepare\_examples.py}, \texttt{scripts/prepare\_model\_viz.py}). The compute figures below and the timings of \cref{sec:speed} come from the training and measurement logs.

\paragraph{Compute.} Data preparation and training ran on one machine with an NVIDIA T4 (16\,GB), 16 vCPUs and 62\,GB of RAM; the training process peaked at about 11\,GB of RAM and 4.6\,GB of GPU memory. The full run took about 5\,h after the datasets were downloaded, of which about 3\,h was training (10 epochs of $\approx$1{,}095\,s). The downloads took about 50\,GB, the audio banks 31.9\,GB and the feature cache 18.0\,GB.

\paragraph{What is needed for exact reproduction.} The released model was warm-started from, and its speech beds and noise pool were cleaned with, a preliminary 31-class checkpoint of the same architecture that was trained earlier on NVS-38K and Vaani (with the same NVS-38K and Vaani split functions) and is not released; the compute figures above exclude its training. The training notebooks use whatever model is in \texttt{checkpoints/} for these two roles, which is now the released model, and the peak learning rate depends on whether a warm-start checkpoint exists. A re-run therefore follows the same procedure but does not reproduce the released weights exactly.

\paragraph{Randomness.} Fixed pools and validation sets are generated with fixed seeds, so they are identical across re-runs; batch order, the time shift and SpecAugment are not seeded. We trained once.

\end{document}